\pdfoutput=1
\documentclass[3p,twocolumn,a4paper,times,12pt]{elsarticle}

\usepackage{graphicx}
\usepackage{makecell}
\usepackage{hyperref}

\newcommand{\idest}{\textit{i.\,e. }}
\newcommand{\eg}{\textit{e.\,g. }}
\newcommand{\HADRON}{HAD\-RON }
\newcommand{\PAMIR}{PA\-MIR }
\newcommand{\KASCADE}{KAS\-CA\-DE }
\newcommand{\ch}[1]{#1}

\begin{document}

\begin{frontmatter}
\title{Scaling violation in interaction of cosmic ray hadrons and the nature of the 3\,PeV~knee in the spectrum of primary cosmic rays}

\author{S.\,Shaulov}
\author{A.\,Shepetov} \ead{ashep@tien-shan.org}
\author{E.\,Kupriyanova}
\author{V.\,Ryabov}
\author{L.\,Vil'danova}

\address{P.\,N.\,Lebedev Physical Institute of the Russian Academy of Sciences (LPI), Leninsky pr., 53, Moscow, Russia, 119991}

\date{}

\begin{abstract}
Energy spectra of the most energetic hadrons in the core region of extensive air showers (EAS) were studied in dependence on the EAS energy $E_0$ in the hybrid experiment HAD\-RON at the Tien Shan mountain cosmic ray station. For the first time by direct measurement it was found in this experiment that the slope of the power energy spectrum of EAS hadrons diminishes itself locally, and the average hadron energy, correspondingly, increases between the $E_0$ borders 3\,PeV and 20\,PeV. Such behavior agrees with threshold appearance in EAS, in the same energy range, of a long-flying penetrative component which was earlier revealed at the Tien Shan station. Now we reconsider this Tien Shan result in comparison with the new data of modern collider experiments. The analysis permits to state unambiguously an astrophysical nature of the penetrative EAS component, and to discuss its probable connection with the origin of the famous knee in the energy spectrum of cosmic rays at $E_0\simeq 3$\,PeV.
\end{abstract}


\end{frontmatter}

%\maketitle

\section{Introduction}

Discovery of a sharp change of power index at the energy $E_0\simeq 3$\,PeV, named the \textit{knee}, in \ch{the} differential energy spectrum ${dI}/{dE_0}\sim E_0^{-\gamma}$ of primary cosmic ray particles \cite{khristiansen} has attracted great interest and caused multitude of experimental and theoretical studies, though until now the origin of the knee still remains under discussion \cite{horandel2004,onknee_haungs2015,horandel2016,onknee2019b,onknee2019}. The only way to solution of the knee puzzle is indirect investigation of cosmic rays by the method of extensive air showers (EAS), \ch{primarily} through determination of the \ch{average mass composition} of cosmic ray particles, and continuous tracing \ch{its variation in the range} of $E_0$ values between $(1-100)$\,PeV. There are two \ch{practical} approaches to do such a study: \ch{an exploration of the electromagnetic component of EAS, which constitutes overwhelming majority of EAS particles}, and precise investigation \ch{only} of the central core in showers  which generally consists of the most energetic hadrons.

Up to the present time \ch{detailed examination} of the data on electromagnetic EAS component has revealed a rather complicated spectrum structure around the knee region. Until the energy of $E_0\simeq 3$\,PeV the index of spectrum slope $\gamma$ there remains equal to $2.75$, then it experiences two breaks at $3$\,PeV and $100$\,PeV, such that $\gamma\simeq 3.0$ between $E_0\simeq (3-100)$\,PeV; then the slope grows up to $3.1-3.3$
%\cite{shaulov_composition_2003b,2005kascade,2008aragats,2012kascade,onspectrum_icetop_2017}.
\cite{shaulov_composition_2009,2005kascade,2008aragats,2012kascade,onspectrum_icetop_2017}. As a rule, by \ch{the} EAS investigation approach any irregularities in the energy spectrum of cosmic rays were \ch{interpreted as discontinuities} in the abundance of particular nuclear components with increasing atomic numbers $Z$, \ch{such that the break for all nuclei types occurs at the same rigidity of magnetic cut-off, $R_{c}\sim E_0/Z$. Above this threshold the deposits of different components into the common spectrum disappear successively, in proportion to their charge $Z$.}

% Up to date, a most abundant statistic in the range of the knee, of about $\sim$$10^8$~of EAS events, has been collected in the \KASCADE and \KASCADE Grand experiments \cite{kascade2013}. According to prevailing interpretation of these data, the leading spectrum knee at $3$\,PeV is connected with extinction of cosmic ray protons, such that the rigidity border $R_{c}$ should be of about $\sim$$3$\,PV, and further on the deposits of heavier nuclei into the common spectrum disappear successively in proportion to their charge $Z$.

Principal shortage of the EAS method is that \ch{interpretation} of raw experimental data is heavily bound to simulation of the development process of nuclear cascades in the air, which, in turn, depends strongly on \ch{supposed} interaction model of high energy particles. This is the reason why there exists considerable discrepancy between the results of EAS experiments which were analyzed with application of different interaction models \ch{ \cite{pao_models_compari_2019,ta_models_compari_2021}}. For example, as it concerns the mass composition of primary cosmic rays at $E_0\sim 1$\,PeV, the share of protons can variate thrice as much between estimations given by various research groups \ch{\cite{composition_review_2020}}.

Essentially more reliable and model-independent information on \ch{the} composition of cosmic ray particles can be fet\-ched from direct investigation of the hadrons \ch{of EAS core}. This method is more complicated since it necessitates building of a solid, large sized detector of the hadronic EAS component.  During a 60\,years long period such facilities were successively realized at the height of 3340\,m above the sea level, at the Tien Shan mountain cosmic ray station of LPI. At the first stage of exploration activity at Tien Shan, in later 1960s--1970s, this was the Large Ionization Calorimeter which had the total absorber depth $850$\,g/cm$^2$, the sensitive area $36$\,m$^2$, and was located in the center of a system of EAS particles detectors.
% \cite{scinti1}.
One of the main results of this experiment was the discovery of the penetrative component of EAS which appears in threshold manner just around the knee of \ch{the} cosmic ray spectrum \cite{ontienlongflyicrc19,longflying1,ontienlongfly2008,ontienlongfly2006}.

The Tien Shan calorimeter was later replaced by a 160\,m$^2$ large X-ray emulsion chamber (XREC). This modification was made in the frames of the experiment \HADRON (1985--1991) which united the X-ray films and EAS detection methods in a single hybrid installation \cite{ontientu,epj2017_shaulov}. Together with essential increase of the size of continuous area of EAS cores detection, the XREC permitted to measure directly the individual energy of the TeV scale gamma ray quanta \ch{inside EAS core}, and to \ch{study their characteristics} in dependence \ch{on the total number of EAS electrons $N_e$ (generally named as EAS \textit{size}), as well as on the energy $E_0$ of the primary cosmic ray particle which had created the EAS}. Since the main source of high-energy gamma rays in EAS core is the radiative decay $\pi^0\rightarrow 2\gamma$, \ch{the spectrum of detected gamma rays carries information on the properties} of the most energetic hadrons (pions) of EAS.

As a result of \HADRON experiment, it was found a non-monotonous local variation of \ch{the power index in the spectra of detected gamma ray quanta, ${dI}/{dE_\gamma}\sim E_\gamma^{-\beta}$. Specific change of energy spectra takes place in the range of primary energies $E_0\simeq (1-100)$\,PeV: with steady increase of the shower size and energy,} the absolute value of $\beta$ index diminishes systematically between the $E_0$ limits of $(3-20)$\,PeV, and then it returns back to its initial, or somewhat larger, value in the interval $E_0\simeq (20-100)$\,PeV.

Any drop in the slope of hadron spectra at observation level is equivalent to \ch{corresponding} rise of \ch{the} average energy of secondary hadrons \ch{which, in turn, reflects reduction of} energy dissipation of \ch{a} primary cosmic ray particle in the atmosphere. Thus, the result of \HADRON experiment confirms the preceding conclusion made at Tien Shan on appearance of the penetrative component around the knee region of the cosmic rays spectrum. An underlying reason of such change in production \ch{of secondary hadrons} may be either variation \ch{in  development} of nuclear cascade because of \ch{some} change in \ch{the} elementary processes of particles interaction at high energies, or appearance \ch{in the flux of cosmic rays of a} lighter, and consequently more penetrating, component, such as protons and He nuclei.

\ch{To date} the technique of high-energy experiments has reached an unique stage, \ch{in the sense that at two} biggest accelerators, the Large Hadron Collider (LHC) and Relativistic Heavy Ion Collider (RHIC), the \ch{invariant center-of-mass} energy of particles collisions  $\sqrt{s}$, if reduced to the target rest frame, is  equivalent to \ch{the} interaction energy slightly below $100$\,PeV, and well above the position of the $3$\,PeV knee in the cosmic ray spectrum. Especially for testing and calibration of hadronic interaction models used for simulation in the cosmic ray physics, \ch{two experiments were carried out} at both colliders, correspondingly LHCf and RHICf (``f'' means ``forward''). \ch{Both experiments were} aimed to detection of neutral products of hadronic reactions, such as the gamma rays from $\pi^0$ decays, neutrons, and neutral kaons \ch{which were} emitted  at small transverse angle relative to the direction of accelerator beam, \idest in typical geometry of a cosmic ray experiment \cite{crvslhc_lhcf_2011,crvslhc_lhcf_2016,crvslhc_rhic_2014,crvslhc_lhcf_2018,crvslhc_uhecr_2018,crvslhc_dembinski_2019}. 

% 7TeV=7000GeV
% 2 (7000^2 GeV^2) / (1 GeV) = 9.8e7 GeV ~ 10^17eV 

In the high energy physics it is common to present the spectra of \ch{interaction} product particles in dependence on the dimensionless Feynman variable $x_F=2p_\|/\sqrt{s}$, \ch{were $p_\|$ is the longitudinal momentum component of a produced particle}
%,  $\sqrt{s}$ is the invariant interaction energy of collision reaction
\cite{feynmanscaling}.
\ch{By investigation of the interactions of cosmic ray particles} with the energy $E_0\gtrsim 1$\,PeV it is \ch{studied} mostly an extremely forward fragmentation region of the kinematic space of reaction products, which corresponds to the values of Feynman variable $x_F\geqslant ( 0.05-0.1)$. In this region the properties of secondary particles depend to the utmost on the momentum and quantum numbers of a \ch{primary} cosmic ray projectile. Well-known hypothesis of limiting fragmentation states that the differential cross-section of product particles, if written in the form
%${x_F}\left/\right.{\sigma_{inel}} \cdot {d\sigma}\left/\right.{dx_{F}}$
$\frac{x_F}{\sigma_{inel}}\cdot\frac{d\sigma}{dx_F}$ and considered as function of $x_F$, should be independent \ch{in the fragmentation region} of interaction energy $\sqrt{s}$ (where $\sigma_{inel}$ is the total cross-section of inelastic interaction) \cite{feynmanscalingchou}. This assertion, \ch{often referenced to as Feynman scaling}, has an universal character and was confirmed in many accelerator experiments \cite{feynmanscaling2}.

\begin{figure*}
{\centering
\includegraphics[width=0.6\textwidth, trim=0mm 0mm 0mm 0mm]{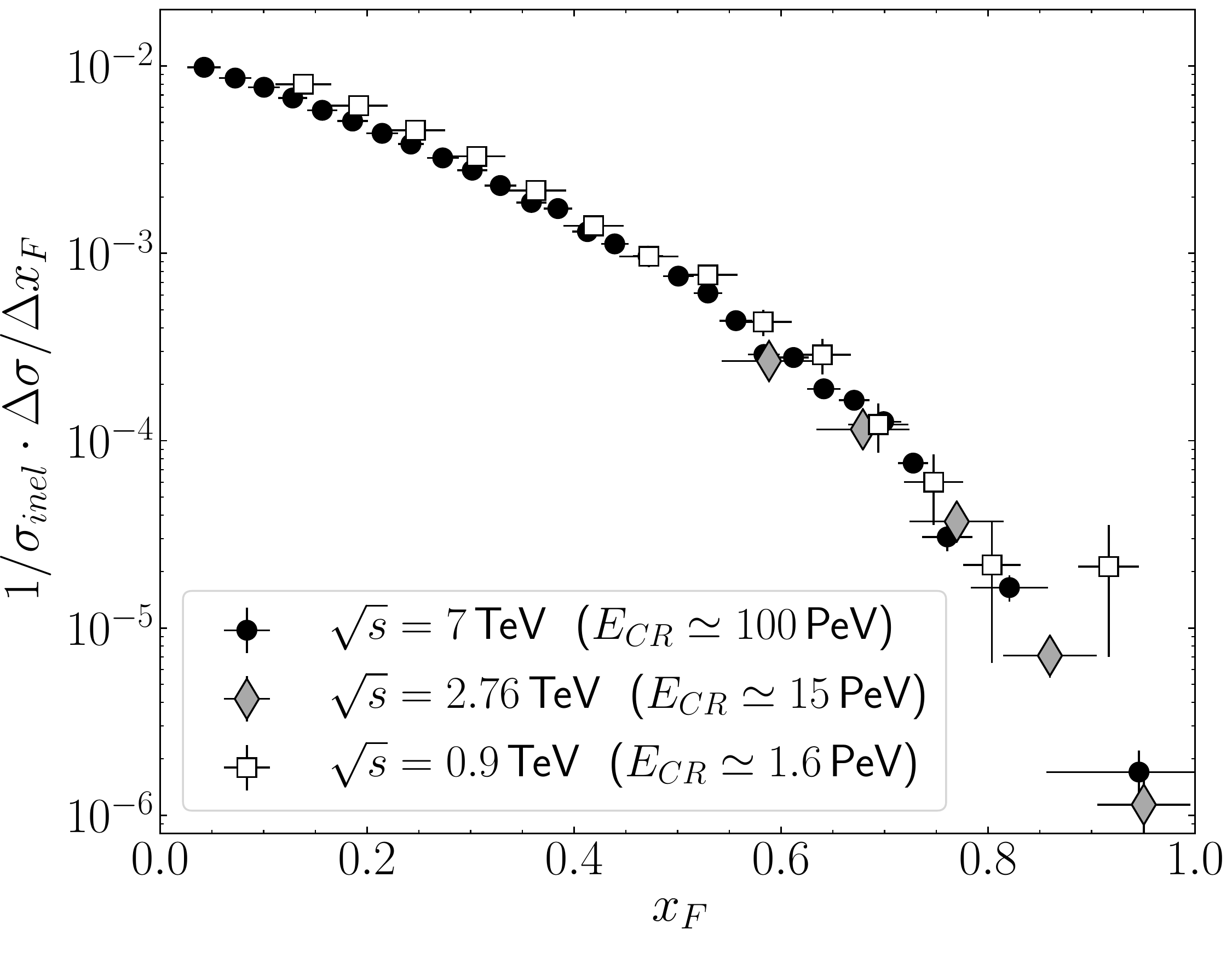}
\caption{\ch{Scaling behavior of the distribution over the Feynman variable of neutral pions produced in $pp$ collisions with the different center of mass energy $\sqrt{s}$ (and with equivalent interaction energy of a cosmic ray proton $E_{CR}$). The original data for the plot were taken from publications \cite{crvslhc_lhcf_2011} and \cite{crvslhc_lhcf_2016}.}}
\label{figiscaling}}
\end{figure*}

Currently the angular and energy spectra of neutrons and gamma rays have been tested in $p-p$ collisions at LHC with invariant energy up to $\sqrt{s}=7$\,TeV, and
% within an extremely forward region of kinematic space of the pseudorapidity range of $|\eta|\geqslant 8.4$ and
under tangential angles $\theta_T\leqslant 0.01^{\circ}$ relative to the beam axis. \ch{As it is illustrated by the plot of Fig.~\ref{figiscaling},} the scaling over the Feynman variable holds indeed among \ch{the distributions of product particles}  which continue to keep their shape untouched independently of the interaction energy $\sqrt{s}$ \cite{crvslhc_lhcf_2011,crvslhc_lhcf_2016}. Neither \ch{any deviation from scaling behavior was found} in collisions of \ch{the heavy nuclei of different types} at the RHIC accelerator \cite{crvslhc_rhic_2014}.

\ch{In terms of the high energy physics studied at colliders the variation found in the \HADRON experiment of the spectrum slope of secondary pions in dependence on the primary EAS energy $E_0$ means an evident scaling violation at interaction of cosmic ray particles. The results newly gained in modern collider experiments permit to exclude any peculiarity of the elementary process of nuclear interaction at $E_0\simeq 3$\,PeV as probable reason for the change in gamma ray spectra detected in the Tien Shan XREC.} Since now on any explanation for this effect must be searched for only in the astrophysical domain, \ch{\idest} among possibilities of some change \ch{in the content of primary cosmic rays}. A modern review of the Tien Shan experimental data with discussion of their possible explanations in the light of the current state of accelerator physics is the subject of present publication.

\section{\ch{The \HADRON experiment}}

\subsection{\ch{The basics of the X-ray emulsion technique}}
\label{sectixrec}

\begin{figure*}
{\centering
\includegraphics[width=0.7\textwidth, trim=0mm 0mm 0mm 0mm]{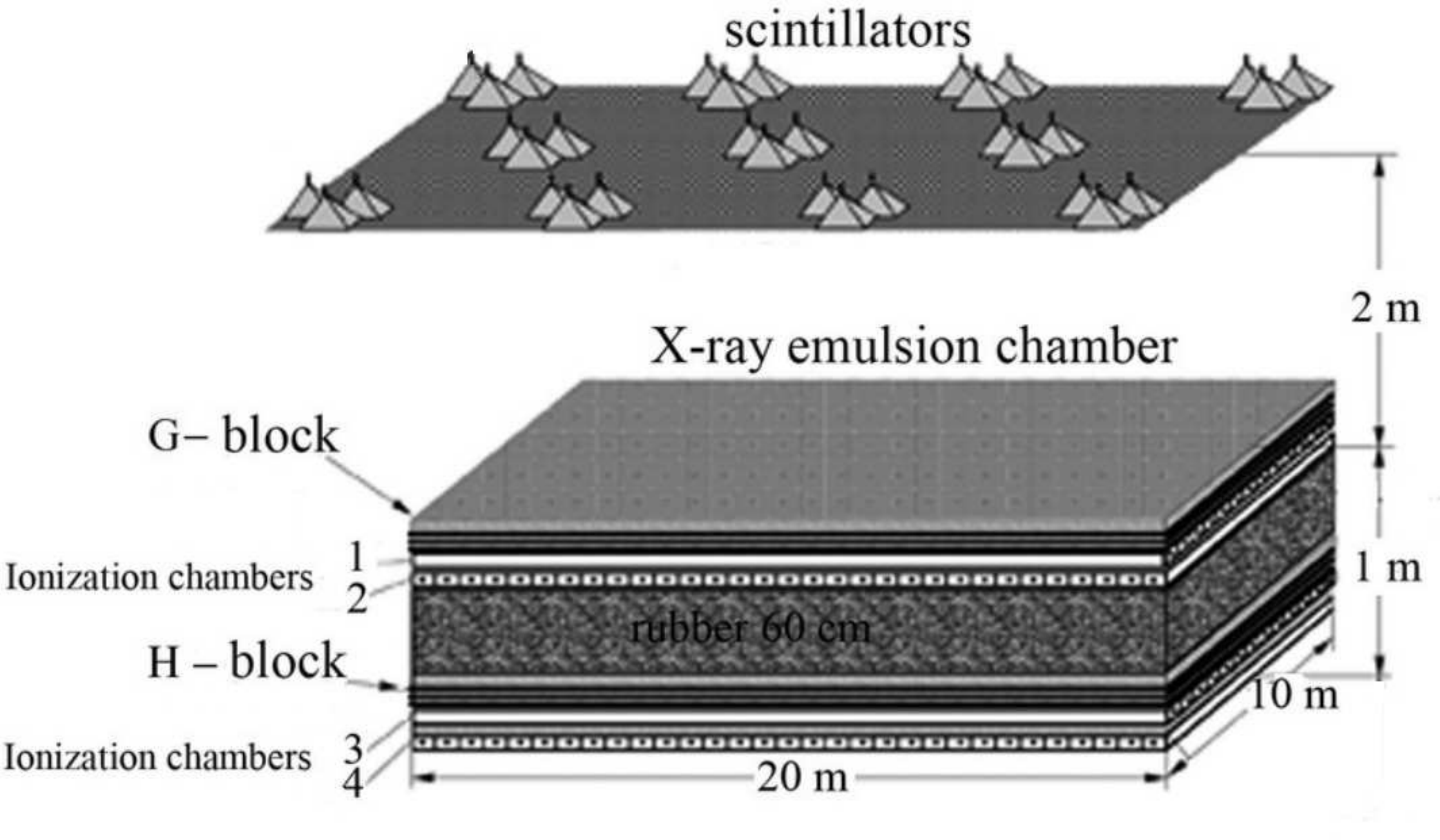}
\caption{\ch{The scheme of detector disposition in the central part of the \HADRON installation. From top to bottom: the ''carpet'' of scintillation detectors for registration of the charged particles of EAS; the detector of the shower core hadrons (X-ray emulsion films and ionization chambers).
}}
\label{figihadro}}
\end{figure*}

For detection of the most energetic particles of EAS cores in the \HADRON experiment it was used an X-ray emulsion chamber (XREC) which had the total sensitive area 160\,m$^2$ and was placed in the center of an installation of EAS \ch{particles} detectors. The chamber was composed by three layers of two-sided X-ray films interlaced by lead plates, so that the total thickness of absorber in the chamber was about $(8-10)$ electromagnetic cascade units\ch{, or about $0.5$ interaction length of high energy hadrons.
The scheme of detector disposition in the center of the \HADRON installation is shown in Fig.\,\ref{figihadro}}.

The operation principle of XREC consists of the following. \ch{When coming into the earth's atmosphere,} a cosmic ray particle with a PeV order energy gives start to \ch{development of an EAS}, in which process deciding role play the high energy hadrons of \ch{shower} core. \ch{Large variety of secondary hadrons is produced in successive interactions both of the primary cosmic ray particle itself, and of its most energetic descendants.  In many simulations made at design epoch of the XREC method it was found that at the energy of primary particle $E_0\gtrsim 100$\,TeV ($\sqrt{s}\gtrsim 1$\,TeV) the main part of interaction products constitute pions. Also, it was demonstrated that the mean production height of the particles arriving to an XREC situated in mountains is $(2-5)$\,km above the installation, and the average amount of successive interactions in the atmosphere occurs about $(2-3)$ \cite{xrec_trudy154}.}
% Семейства гамма-квантов ТэВ ных энергий генерируются  примерно в 2-3 взаимодействиях на высоте 2-4 км над установкой.

\ch{
In particular, one third of total multitude of pion particles produced in a high energy hadronic cascade make up the neutral pions of a TeV scale energy, which practically momentarily decay by radiative channel $\pi^0\rightarrow 2\gamma$ (with a mean lifetime $\tau\sim 10^{-16}$\,s only in their rest frame). When hitting the XREC, the high-energy gamma rays from $\pi^0$ decays which have taken place in the air above the installation initiate electron-photon cascades within the dense lead absorber, with the total amount of electrons at the maximum of cascade development depending on the energy of initial gamma quantum. In an electromagnetic cascade originating from a TeV energy quantum and developing in lead the peak density of electrons is sufficient to cause darkening of X-ray emulsion, such that the value of particles density at cascade maximum, as well as the energy of the initial quantum and of its parent pion can be estimated through photo-densitometry of the darkened spot in the film
\cite{xrec_trudy154,xrec_civilizedanalog154,onxrec_grundlagen_1971}.
}

\ch{
In the considered range of cosmic ray particles with primary energy $E_0\gtrsim 100$\,TeV the total cross-section of purely hadronic interaction, $\sigma_{hh}\geqslant 80$\,mb, is $\sim$$1000$ times} \ch{as large as the production cross-section of gamma rays, $\sigma_{h\gamma }\sim 0.1$\,mb. (The latter are mostly produced through creation of $e^{\pm}$ pairs in the atmospheric cascade, with subsequent annihilation of positrons). Thus, the described origin of the gamma ray quanta detected in XREC from decay of neutral pions resolutely prevails above any other sources of high energy gamma rays.

Presently, the multiple production of high-energy gamma rays and other neutral products by interaction of the PeV scale hadrons (in the laboratory frame) has been immediately, and quite independently, confirmed by modern collider experiments, such as LHCf and RHICf mentioned in Introduction section.
}

\ch{
As for the relatively long-living charged pions ($\tau\gtrsim 10^{-8}$\,s in the rest frame) and baryons of EAS core, they can reach the installation level and interact within the internal material of XREC, which results, again, in development of an electromagnetic cascade in the lead absorber. Since the total thickness of the Tien Shan XREC was not well enough for effective absorption of high energy hadrons, in a part of the whole exposition an additional $60$\,cm thick target of carbon enriched rubber ((CH)$_n$) was placed between the two similar XREC units stacked one above the other \cite{ontientu,epj2017_shaulov}. Thus, the electromagnetic cascades in the upper unit ($\Gamma$-block) were mostly generated by the gamma rays from high energy $\pi^0$ decays in the atmosphere, while the neutral pions from hadronic interactions in the rubber target were detected in analogous manner in the lower unit (H-block).
}

\ch{
Further on throughout this article the review of the \HADRON experiment is made on the basis of the data obtained only in the upper $\Gamma$-block of the Tien Shan XREC. Generally, the gamma ray quanta of a TeV scale energy, just like their parent pions, are concentrated in the central region of developing EAS, so that the Tien Shan detector \HADRON was a specialized tool for precise investigation of the structure of high-energy hadronic core in extensive air showers.
}

%The minimum energy threshold of cosmic ray hadrons detection at the Tien Shan XREC was of about $(1-2) / K_\gamma$\,TeV, where $K_\gamma\simeq (0.15-0.2)$ is the average coefficient of energy transfer into the electromagnetic component at hadronic interactions. Generally, the hadrons of such energy concentrate in the central region of developing EAS, so the Tien Shan detector \HADRON was a specialized tool for precise investigation of the structure of high-energy hadronic core in extensive air showers.

Of no small importance is the fact that estimation \ch{of the energy of initial gamma} \ch{ray quantum} by the XREC method is exclusively based on \ch{photo-densitometry of a darkening spot which has been leaved in the X-ray film by an electron-photon cascade developing in lead. The latter is a purely electromagnetic process accessible for complete mathematical calculation,} and does not depend on any specific model of hadronic interaction. It can be stated \ch{therefore} that the XREC method is a mean for direct, \ch{\idest model-independent,} investigation of the properties of high-energy gamma rays, and consequently of their parent hadrons in EAS core.

\ch{
Besides the energy of gamma ray quanta, an information was also available on their direction angles, both zenith $\theta$ and azimuth $\phi$, thanks to using in the Tien Shan XREC two-sided films with an emulsion layer on each side. The direction angles of electromagnetic cascade in this case were defined by mutual displacement of the spots in two emulsion layers separated by a 200\,$\mu$m thick plastic substrate. The mean estimation errors of the cascade angles in the Tien Shan XREC were $\Delta\theta\approx 3^\circ$ and $\Delta\phi\approx 15^\circ$.
}

\begin{figure*}
{\centering
\includegraphics[width=0.8\textwidth, trim=0mm 0mm 0mm 0mm]{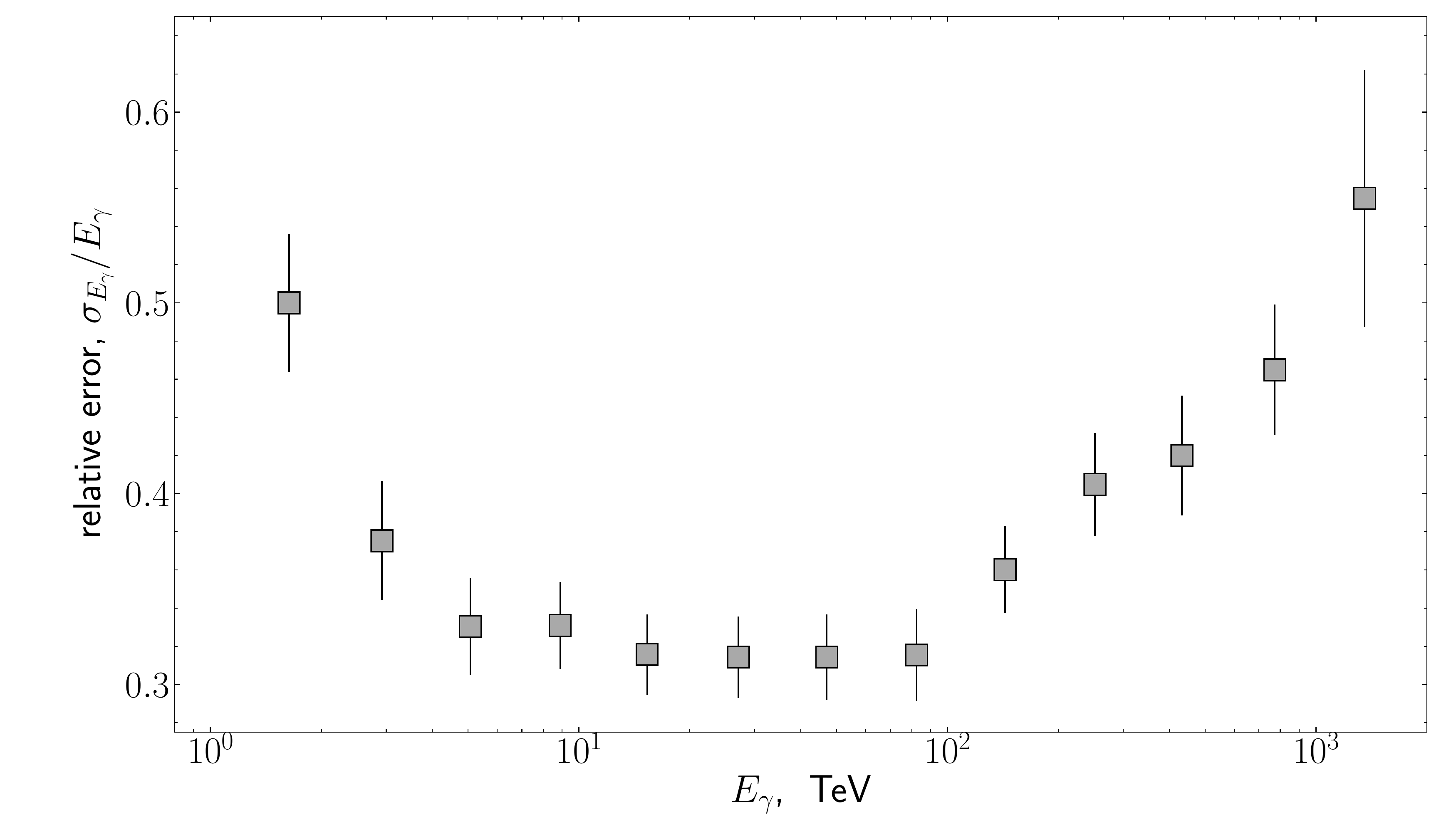}
\caption{The average error of energy estimation by the XREC method.}
\label{figishaulov_xrecerrs}}
\end{figure*}

\ch{
Calibration of the method of particles energy determination by the photo-densitometry of X-ray films was made in special accelerator experiments in the beams of electrons and pions. In cosmic rays the accuracy of energy estimates was tested in an experiment on reconstruction of the invariant mass of neutral pion by measuring, by the means of XREC, the energies $E_{\gamma_1}$, $E_{\gamma_2}$, and opening angle $\varphi$ in the pairs of gamma ray quanta originating from $\pi^0$ decays: $m_{\pi^0}=\varphi\cdot\sqrt{E_{\gamma_1}\cdot E_{\gamma_2}}$  \cite{xrec_civilizedanalog154_on_pi0_calibration,onxrec_pi0calibri}.} As shown in Fig.~\ref{figishaulov_xrecerrs}, in a wide range of gamma ray energy $E_\gamma$ the precision of the method remains of about $(30-40)$\%.

\ch{
Since the lateral size of the darkening spot from electromagnetic cascade in X-ray film is of a few tens of micrometers only, a} characteristic feature of the XREC technique \ch{applied for cosmic ray study} is its unprecedentedly high spatial resolution, typically $\lesssim$$100$\,$\mu$m, which permits to register the electron-photon cascades \ch{created} by particular gamma ray quanta even \ch{in the very center of high-energy EAS}. In the Tien Shan experiment \HADRON an excellent spatial resolution and acceptable precision of energy measurement \ch{of the individual particles in EAS cores}, ensured by the XREC, were combined with estimation possibility of the energy of primary \ch{interaction} $E_0$ by information from the Tien Shan system of EAS detectors.  This combination made the \HADRON installation an unique instrument for detailed investigation of the properties of hadronic interactions in the $(1-100)$\,PeV energy range (in the laboratory frame).

In particular, the spatial resolution of X-ray films permitted to detect separately the electron-photon cascades \ch{left by the} tight \ch{bunches} of genetically connected high-energy gamma rays which had emerged from interaction of the same cosmic ray particle. In \ch{the} scientific literature on XREC \ch{visible traces of such interactions} are commonly referenced to as \textit{families}. {As a rule, a gamma family reveals itself as a group of the darkening spots with coinciding direction angles $(\theta,\phi)$ distributed within a limited space with typical diameter of $\sim$30\,cm on the film.}

The main goal of the \HADRON experiment was investigation, \ch{in dependence on the primary interaction energy $E_0$, of the properties of gamma families in the Tien Shan XREC.} Among those properties there were the \ch{multiplicity $N_\gamma$ and energy $E_\gamma$ of individual gamma ray quanta in a family, the energy spectrum $d N_\gamma / d E_\gamma$ and sum energy $\sum E_{\gamma}$ of gamma ray quanta in families, total multiplicity of the families $N_{fam}$, and the relative share of the EASs accompanied with families in the XREC, $N_{fam}/N_{EAS}$. As well, it was of interest the dependence of all these characteristics on the parameters of corresponding shower, first of all on its size (the total number of shower particles) $N_e$ and the energy $E_0$ of a primary cosmic ray particle which has caused development of the EAS.}

\subsection{\ch{Detection of EAS particles}}
\label{sectieas}

\ch{
The parameters of EAS in the \HADRON experiment were derived from the information on distribution of the charged particles density  in shower. The latter was registered by a system of $\sim$$100$ scintillation detectors installed both immediately above the XREC top ($11$~points of the central detector ''carpet'', see  Fig.\,\ref{figihadro}), and along the circles with $20$\,m, $50$\,m, and $80$\,m radii circumscribed around this center. All detectors were made of the $(0.5\times 0.5\times 0.05)$\,m$^3$ blocks of solid polystyrene scintillator \cite{epj2017_shaulov,ontienhistory1,ontien_icrc1987___through_ne_ru}. In the case of an EAS event the sum amplitude of scintillation flash generated in such detector is proportional to the amount of charged particles which have passed through the point of detector location, so registration of those amplitudes in a system of synchronously operating detectors gives the spatial distribution of particles density in a shower.
}

\ch{
During the measurements the whole system of particles detectors operated under the control of a common trigger signal which was generated in the cases when the total amplitude of scintillation pulses summed over all detectors exceeded a predefined threshold. The settings of the trigger elaboration scheme in the time of \HADRON experiment ensured effective detection of all EAS with the primary energy $E_0\geqslant 0.2$\,PeV whose axes were passing through the area of the central ''carpet''.
}

\ch{
Further processing of the particles density data registered by EAS passage consisted} \ch{of determination of the basic characteristics of EAS: the $(x, y)$ coordinates pair of the shower axis location, the ''age'' parameter $s$ of the shower, and an estimate for the total number of particles $N_e$ (the EAS ''size''). This was made by fitting of the two-dimensional distribution of experimentally measured particles density $\rho_{exp}$ with a function of the Nishimura-Kamata-Greisen family (NKG, \cite{nkg-nk}):
}
$$
\rho_{NKG}( r, s, N_e )= 
0.366 s^2 ( 2.07-s )^{1.25}\times
$$
$$
( r / r_M )^{s - 2 }( 1 + r / r_M )^{s - 4.5} / r_M^2 )\times
N_e,
$$
$$
\sum\frac{(\rho_{NKG}( r(x,y), s, N_e )-\rho_{exp})^2}{\sigma_\rho^2}\rightarrow \min_{x,y,s,N_e},
$$
where the parameter of Moli\'{e}r radius $r_M$ equals to 120\,m at the altitude of the Tien~Shan station.
Average measurement errors of the particles density in the detector points $\sigma_\rho$, as well as typical uncertainties of the shower parameters estimation by the said algorithm were deduced from simulations of operation of the detectors system. It was found by simulation that the relative accuracy $\Delta N_e/N_e$ of the EAS size estimates is about $(5-10)$\%, while the mean quadratic error of axis position equals to $(3-5)$\,m.

%Of them, the work \cite{2007aragats_simul_like_tien} could be mentioned which was made using contemporary models of particles interaction, such as QGSJET, for the EAS installation \GAMMA at Aragats. The latter was a total analog of the Tien Shan experiment, build on the particles detector devices of the same type, and situated at the same height above the sea level. This simulation has shown that the relative accuracy $\Delta N_e/N_e$ of EAS size parameters restored by the density distribution of shower particles among the detectors of such installation is of about $(5-10)$\%, while the mean quadratic error of the EAS axis position equals to $(3-5)$\,m. 

\ch{
By the fitting procedure the zenith angle of shower axis $\theta$ was also taken into account which was defined, together with the azimuth angle $\phi$, by the time delays between the fronts of scintillation flash in several spatially separated points of the detector system. The typical estimation errors of the direction angles of EAS by this method were $\Delta\theta\approx 8^\circ$ and $\Delta\phi\approx 13^\circ$.
}

\ch{
Thus, the estimation method of shower parameters applied in the \HADRON experiment was based exclusively on experimental measurement of the particles density, and did not depend on any specific model of EAS development.
}

\ch{
An alternative, and completely independent way to locate the position of EAS center in the \HADRON experiment was using the amplitude distribution of electromagnetic signal among the two layers of ionization chambers with mutually perpendicular long axes which covered the whole area of the Tien Shan XREC. The chambers constituted a 2D coordinate frame which permitted to locate EAS core on the surface of the XREC with accuracy $\pm$$0.25$\,m, an order of magnitude better than positioning of EAS axis by the data of the scintillation detectors system. This was especially important for the matching task of gamma families with corresponding EAS events (see below).
}

\ch{
With the known size parameter $N_e$ of EAS the energy of a primary cosmic ray particle $E_0$ which has caused development of the shower can be evaluated as $E_0\simeq (2.5\cdot 10^{-6} \times N_e)$\,PeV, with an average relative error of $(25-30)$\%.  This relation was deduced as generalization of simulations of EAS development which have been formerly made for the Tien Shan installation, \eg \cite{tamada_simul_1994_e0_from_ne}. As based on simulation, any estimates of the primary energy $E_0$ in the \HADRON experiment were heavily model-dependent, in contrast to the initial $N_e$ data. This is the reason why any discussion of the results of \HADRON experiment throughout this article will be preferably made in terms of EAS sizes $N_e$, instead of energy $E_0$.
}

I\ch{n particular, from the above formula for estimation of the primary EAS energy it follows that the value $N_e\simeq 10^6$ corresponds to the showers originating from the cosmic ray particles which belong to the $E_0\simeq 3$\,PeV knee region in the energy spectrum of cosmic rays. As well, all the $E_0$ estimates mentioned hereafter in the paper were defined in accordance with this re-calculation rule.
}

\subsection{\ch{Matching the EAS with corresponding XREC families}}

\ch{
Since the duration of continuous exposition of X-ray films in the Tien Shan XREC was of about $0.5$--$1$~year, the main difficulty by selection of an EAS event which could correspond to some particular gamma family is absence of any definite information on the time  of family registration, such that correlation of the EAS and XREC datasets can be made by statistical methods only. Mathematically this is a problem of formal selection between the two alternative hypotheses for every pair of the $i$-th EAS and $j$-th gamma family: $H_0$, \idest that the events do \textit{not} correspond to each other, and $H_1$ if otherwise. The probability of possible error which could be committed by such selection can be represented as combination of two constituents: the false positive probability $\alpha$ to prefer the correlation in the case of actually independent events $i$ and $j$, \idest to accept the hypothesis $H_1$ instead of correct $H_0$, and the false negative probability $\beta$ to reject erroneously the hypothesis $H_1$ in the case of false $H_0$. As known, an appropriate method to deal with such a kind of statistical problems is the Neyman-Pearson likelihood criterion which strives to maximize the probability of \textit{true positive} selection
}
$$
1-\beta=\int_{\omega} f(\bar{x}|H_1)d\bar{x}
$$
\ch{
for any fixed level of the false positive one:
}
$$
\alpha=\int_{\omega} f(\bar{x}|H_0)d\bar{x}.
$$
\ch{
Here, $f(\bar{x}|H_1)$ and $f(\bar{x}|H_0)$ are the conditional probability density functions defined in a common space $\bar{x}=(x_1, \ldots, x_n)$ of the $i$-th and $j$-th events parameters, correspondingly for the truth cases of the hypothesis $H_1$ and $H_0$, and integration is made over a subspace $\omega$ of the $\bar{x}$ points which satisfy the condition
}
$$
\ell(\bar{x})\equiv \frac{f(\bar{x}|H_1)}{f(\bar{x}|H_0)}\geqslant C.
$$

\ch{
The Neyman-Pearson criterion states that in any case when the value of the likelihood function $\ell(\bar{x})$ occurs above some constant threshold $C$ the $H_1$ hypothesis should be accepted, \idest both the EAS and family events \textit{do} correspond to each other. In the opposite case of $\ell(\bar{x})< C$ the hypothesis $H_0$ must be preferred.
The threshold constant $C$ for using in comparison of the likelihood function should be defined experimentally for any  pre-set admissible share of incorrectly matched (false positive) EAS and gamma families. These erroneous correlations constituted  background of} \ch{randomly coinciding families and EAS which was principally unavoidable in the \HADRON experiment because of statistical nature of the correlation procedure.
}

\ch{
For the correlation task of the EAS and XREC events the minimum  set of essential parameters included into calculation of the likelihood function consists of the two projection pairs of zenith angle to the coordinate axes in the $i$-th EAS and $j$-th gamma family, $(\theta_x^i,\theta_y^i)$ and $(\theta_x^j,\theta_y^j)$, of the vector of spatial coordinates of the EAS axis, $\bar{r}^i$, and of the coordinates of the mean energy weighted center of the family $\bar{r}^j$ (the latter was defined as $\bar{r}^j=\sum{E_\gamma^j}\cdot \bar{r}_\gamma^j/\sum E_\gamma^j$ with summation made over all gamma quanta of the $j$-th family). Using the $(\theta_x, \theta_y)$ projection pairs instead of the angles $(\theta,\phi)$ is stipulated by the circumstance that the latter two have a non-Gaussian error distribution. The search rule for correlation between the families and EAS in that case can be written as
}
$$
\ell^{ij}= \ell_{r}^{ij}\cdot \ell_{\theta_x}^{ij}\cdot \ell_{\theta_y}^{ij} \geqslant C,
$$
$$
\mathcal{L}^{j}=\max_{i}{\ell^{ij}},
$$
\ch{
where for every $j$-th family the maximum value of likelihood function was searched for among the shower events which satisfy the threshold condition $\ell^{ij} \geqslant C$. The EAS with the highest $\mathcal{L}^{j}$ estimate was accepted as a most probable counterpart of the $j$-th family. Strictly speaking, the necessity to search additionally for the event with a maximum likelihood score among a set of possible candidates (which were often not unique in practice) means some modification of the classic Neyman-Pearson criterion.
}

\ch{
With account of conditional probabilities $f(\bar{x}|H_{0})$  and $f(\bar{x}|H_{1})$ in the general form of the Neyman-Pearson criterion, an explicit expression for the spatial and angular terms of the likelihood function for using in the \HADRON experiment was written as the following:
}
$$
\ell_{r}^{ij}=\frac{S_{XREC}}{\Delta S_{EAS}} \cdot \frac{1}{2\pi \cdot \sigma_{r^{EAS}} \cdot \sigma_{r^{XREC}}}\times
$$
$$ 
\exp \left(-\frac{ ( \bar{r}^{i} - \bar{r}^{j} )^2}{2\cdot( \sigma_{r^{EAS}}^2 + \sigma_{r^{XREC}}^2)}\right),
$$

$$
\ell_{\theta_x}^{ij}=\frac{P(\theta_x)}{P^{EAS}(\theta_x^{i})\cdot P^{XREC}(\theta_x^{j})} \cdot \frac{1}{2\pi\cdot \sigma_{\theta^{EAS}}\cdot \sigma_{\theta^{XREC}}}\times
$$
$$
\exp \left(-\frac{ ( \theta_x^{i} - \theta_x^{j} )^2}{2\cdot( \sigma_{\theta^{EAS}}^2 + \sigma_{\theta^{XREC}}^2)}\right),
$$
\ch{
and analogically for $\ell_{\theta_y}^{ij}$.
}

\ch{
Here, $S_{XREC}=160$\,m$^2$ is the total area of the Tien Shan XREC, $\Delta S_{EAS}=0.25\times 0.25$\,m$^2$ is the determination accuracy of EAS axis in the intersecting rows of ionization chambers (see Section\,\ref{sectieas}), $P$ is the probability distribution of the actual zenith angles projections in showers, and $P^{EAS}$ and $P^{XREC}$ correspond to the distributions of their estimates made by the means of the methods considered in Sections\,\ref{sectixrec} and\,\ref{sectieas}; $\sigma_{r^{EAS}}$, $\sigma_{\theta^{EAS}}$, $\sigma_{r^{XREC}}$, and $\sigma_{\theta^{XREC}}$ are the estimation errors of the corresponding parameters.
}

%%%%%%%%%%%%%%%%%%%%%%%%%%%%%%%%%%%%%
\ch{
An account to the background of random coincidences between the XREC and EAS events in the \HADRON experiment can be deduced from an analysis of a pair of statistical distributions, $H_0$ and $H_1$, of the highest likelihood estimates $\mathcal{L}$ each of which relates to the truth cases of corresponding hypothesis.
}

I\ch{n practice the distribution $H_0$ (false positive matching) can be explicitly defined by ascribing the random axis coordinates to all EAS and executing afterwards in usual way the correlation procedure between the gamma families and displaced showers. The distribution $H_1$ (true positive matching) is unknown, and only some intermediate distribution of the $\mathcal{L}$ parameter can be experimentally obtained over a set of the correctly correlated EAS--family pairs mixed with randomly coinciding events. The pure appearance of the distribution $H_1$  can be received only in model simulations, when the genetic connection between all EAS and families is apriori known.
}

\ch{
By correlation attempt of every $j$-th family with EAS events there can be met three possibilities: either the highest likelihood estimate remains below the $C$ threshold of the Neyman-Pearson criterion for all shower candidates, $\mathcal{L}^j<C$, or it exceeds the threshold, $\mathcal{L}^j\geqslant C$, or at least any proper EAS counterpart is absent at all (\eg because of the shower detectors system being off at the moment of the family registration). Further on, the relative shares of these alternatives will be designated as $a_1$, $a_2$, and $a_3$ for the experimental gamma families, and as $b_1$, $b_2$, and $b_3$ for the background events. Also, the next designations will be used below: $m$---the probability for some background $b$-th EAS with a likelihood estimate $\ell^{bj}\geqslant C$ to supersede a truly corresponding $t$-th shower because of a larger likelihood, $\ell^{bj}\geqslant \ell^{tj}$, and $\eta$---the relation of the operation time of electronic part of the experimental installation to the whole duration of the XREC exposition.
}

\ch{
Together with the probabilities $\alpha$ and $\beta$ of  false positive and false negative events matching defined above, all the parameters mentioned here are interconnected by the following system of equations:
}
\begin{equation}
\left\{
\begin{array}{l}
a_1=(1 -b_2)\alpha\eta+b_1(1 -\eta), \\
a_2=\underbrace{(1 -m)\beta(1 -\alpha)\eta+(1 -\beta)(1 -\alpha)\eta}_{\mbox{true matching}}+\\
\begin{array}{ccccc}
&&&&
\underbrace{b_2\alpha\eta+m\beta(1 -\alpha)\eta+b_2(1 -\eta)}_{
\mbox{background}},
\end{array}
\\
a_3=b_3(1 -\eta),\\
b_1=(1 -\beta)(1 -b_3),\\
b_2=\beta(1 -b_3),\\
a_1+a_2+a_3=1,\\
b_1+b_2+b_3=1.\\
\end{array}
\right.
\end{equation}
\ch{
Also, these parameters can be helpful for calculation of the amount $T$ of truly matched XREC--EAS event pairs, and of the amount $B$ of the background events resulting from random coincidences:
}
$$
T=\eta(1 -\alpha)(1 -m\beta),
$$
$$
B=b_2\alpha\eta+m\beta\eta(1 -\alpha)+b_2(1 -\eta).
$$
\ch{
Then, the relative share of background among the matched events is
}
$$
\Delta_b = \frac{B}{B+T}=\frac{1+m\beta{f}}{1+f},
\mbox{ where }
f=\frac{\eta(1 -\alpha)}{b_2(1 -\eta(1 -\alpha))}.
$$
\ch{
After excluding the unknown probabilities $\alpha$ and $\beta$ with the help of the above equations system the next expression can be written for the share of background:
}
$$
\Delta_b=\frac{b_2}{1 -b_2}\frac{1 -a_2}{a_2}\left(1+m\frac{a_2 -
b_2}{(1 -b_3)(1 -a_2)}\right).
$$
\ch{
With exception of $m$ all the parameters here can be obtained immediately from the experimental data. The probability $m$ must be defined by simulation of artificial events.
}

\begin{figure*}
{\centering
\includegraphics[width=0.8\textwidth, trim=0mm 0mm 0mm 0mm]{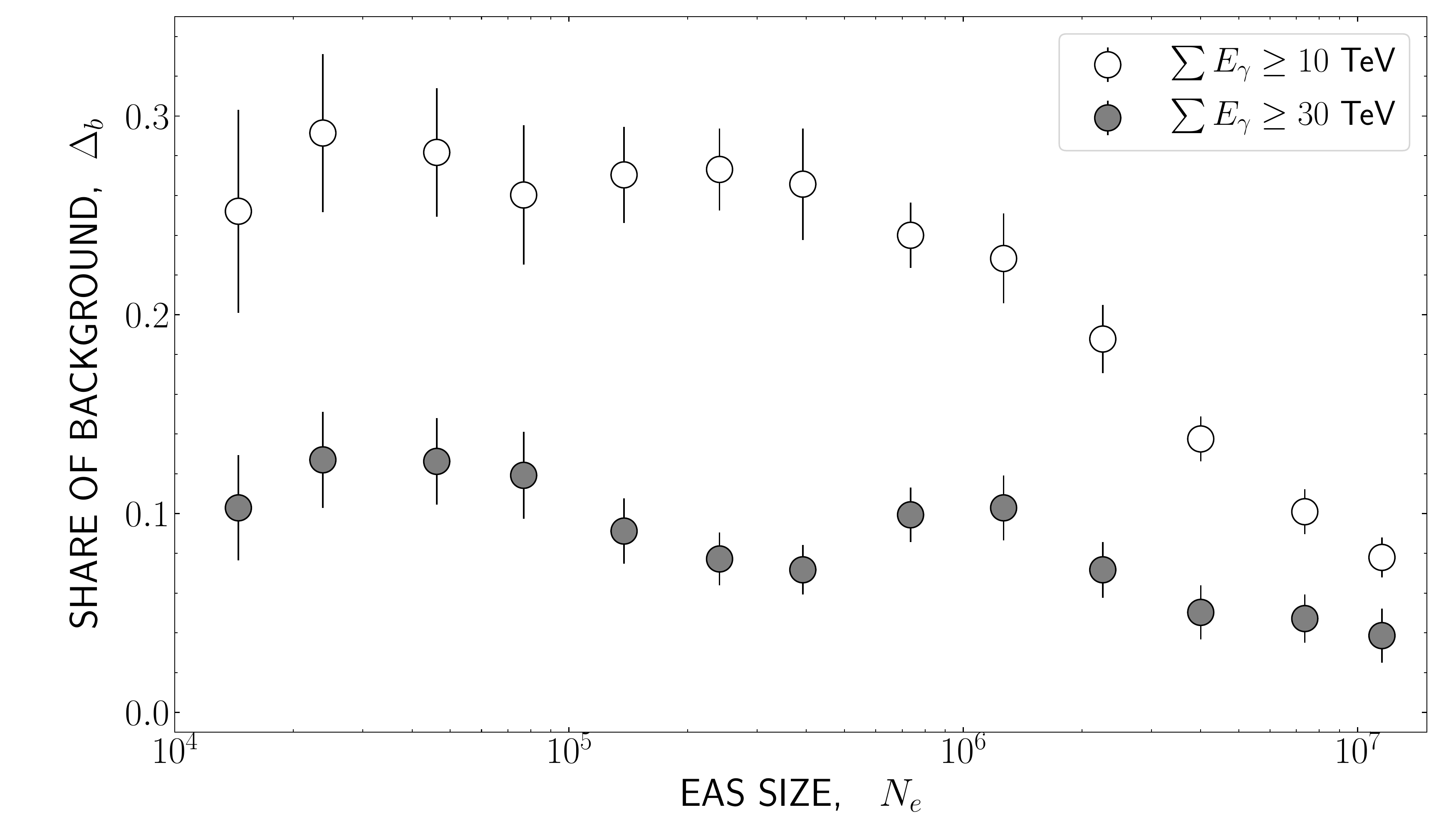}
\caption{\ch{The relative share of the falsely positive background events admitted by the matching procedure of the EAS with XREC families in the \HADRON experiment, in dependence on the size $N_e$ of candidate EAS and on the minimum energy threshold $\sum E_\gamma$ of considered families.}}
\label{figishaulov_sopo}}
\end{figure*}

\ch{
Such simulation was made for the Tien Shan XREC, and corresponding quantitative estimations of the background share $\Delta_b$ were obtained as its result for the various rules of family selection. For the families chosen with the two threshold conditions on the minimum sum energy of gamma ray quanta in the family, $\sum E_\gamma \geqslant 10$\,TeV and  $\sum E_\gamma \geqslant 30$\,TeV, the dependencies of the fraction of incorrectly matched EAS on the average size $N_e$ of candidate showers are shown in Fig.\,\ref{figishaulov_sopo}. It is seen  that among the showers with sizes $N_e\gtrsim 10^5$ this fraction is about $\sim$$(20-25)$\% and $\sim$$10$\%, correspondingly for the two said energy thresholds. In practice the mean energy of gamma families in the Tien Shan XREC was essentially above $30$\,TeV (\eg see Fig.\,\ref{figitamada_nf_vs_ne} below), so the percentage of incorrectly matched events must be even lesser, and the $\Delta_b\simeq 10$\% estimate deduced here should be rather considered in the sense of upper limit for the possible background share of the EAS and XREC events erroneously correlated in the \HADRON experiment.
}

\ch{
By operation of the X-ray films after their development initial selection of the XREC families for further analysis was made in accordance with the next, rather liberal conditions: $E_\gamma\geqslant 2$\,TeV, $N_\gamma\geqslant 1$, $\sum E_{\gamma} \geqslant 10$\,TeV. (Here, $E_\gamma$ is the energy of an individual gamma ray quantum, $\sum E_\gamma$ and $N_\gamma$ are the sum energy and multiplicity of} \ch{the quanta in a family). Among EAS the showers with the size $N_e\geqslant 10^5$ ($E_0\geqslant 0.2$\,PeV) were considered whose axes were passing through the surface of the XREC. During the four expositions of X-ray films made in 1985--1988 in the frames of the \HADRON experiment  it was found $1665$ pairs of events consisting of a gamma ray family in the $G$-block of the Tien Shan XREC and of corresponding EAS. The total statistic of showers which satisfy the said selection criteria and were registered at those times amounts to $3.3\cdot 10^5$.
}

\section{The properties of the hadronic component in the cores of $(1-100)$\,PeV EAS}

\subsection{Differential energy spectra of gamma ray quanta in XREC families}

\begin{figure*}
{\centering
\includegraphics[width=0.57\textwidth, trim=0mm 0mm 0mm 0mm]{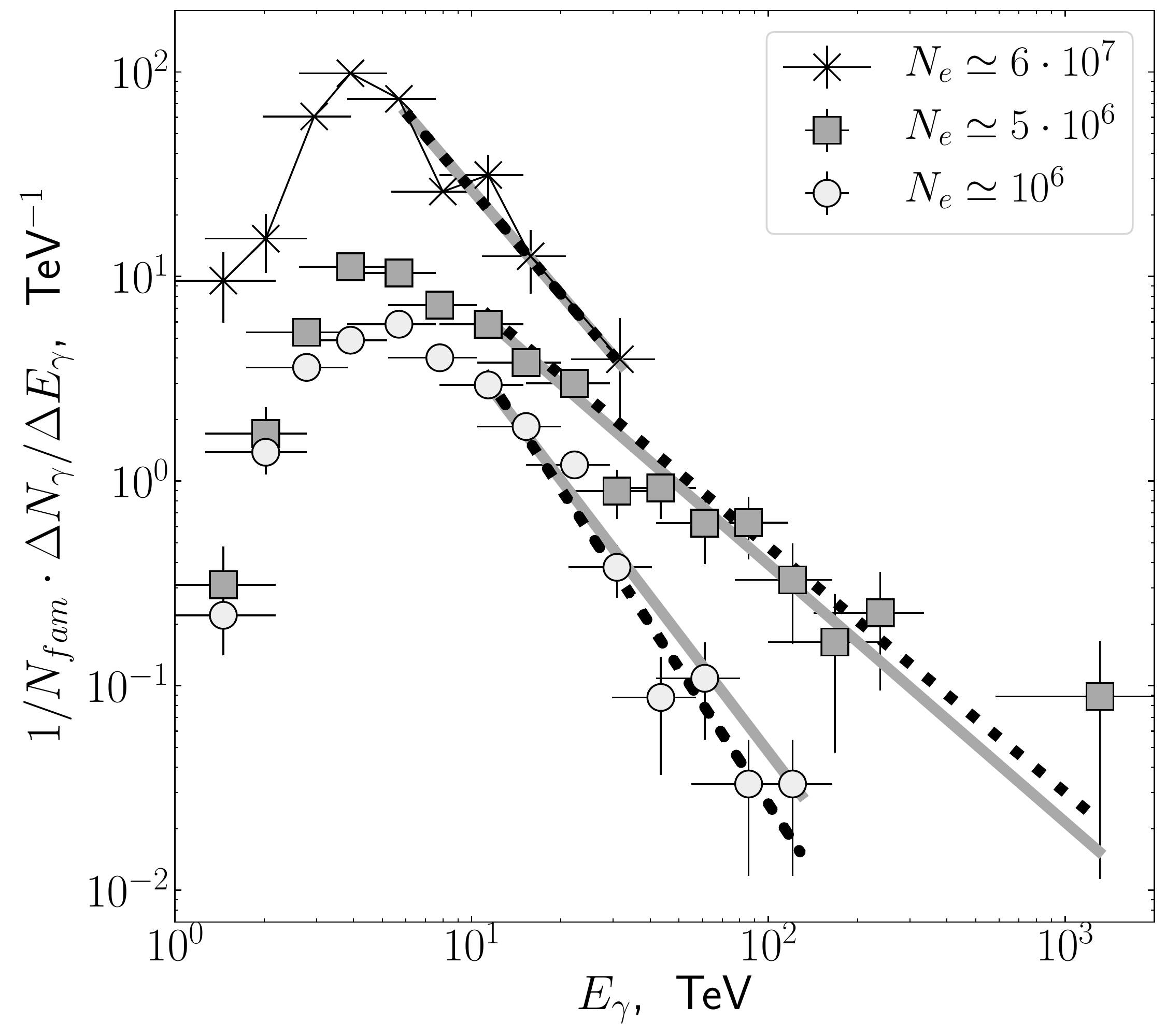}
\caption{\ch{Energy spectra $\Delta N_\gamma/\Delta E_\gamma$ of  gamma ray quanta in the families of the \HADRON experiment (points) averaged between the EAS events with close shower sizes $N_e$. The distributions} are normalized to the total amount of families $N_{fam}$ in each ${N_e}$ group. \ch{The horizontal error bars are marked in compliance with the $E_\gamma$ estimation errors from Fig.~\ref{figishaulov_xrecerrs}.} Straight lines correspond to the best fit power approximation $dN_\gamma/d E_\gamma \sim E_\gamma^{-\beta}$ \ch{of the falling part of energy distributions} defined by the two different methods (see text).
}
\label{figishaulov_spcd}}
\end{figure*}

As a starting point for an analysis of the results of \HADRON experiment it can be used variation of \ch{the} energy spectrum of gamma rays with growth of the size of accompanying shower $N_e$. A sample of such data is presented in Fig.~\ref{figishaulov_spcd} where the differential gamma ray spectra
%$\Delta N_\gamma/\Delta E_\gamma$
are plotted \ch{which were detected in the families with various size values $N_e$ of corresponding EAS}. The distributions presented here were calculated as \ch{an average} between the events which belong to \ch{the three} groups of showers with close $N_e$, such that the mean size for the first group is $\overline{N_e}=10^6$, then the groups with $\overline{N_e}=5\cdot 10^6$ and $\overline{N_e}=6\cdot 10^7$ follow. After calculation each spectrum was normalized to the number of families $N_{fam}$ in corresponding group.

As it can be seen in Fig.~\ref{figishaulov_spcd}, after an initial part where the \ch{shape of gamma ray spectrum is distorted} by the threshold of XREC registration, all distributions \ch{restore the regular power form} near the same energy of gamma ray quanta, $E_\gamma\simeq (6-8)$\,TeV. The further \ch{behavior} of their decreasing tails \ch{is} essentially different: the gamma ray spectrum drawn for the families which belong to the middle $N_e$ range, the one with the average EAS size at $\overline{N_e}=(5\cdot 10^6)$, \ch{drops} down much \ch{slower} then the other two. As well, this is the only distribution in which the gamma ray quanta with anomalously high energy of $E_\gamma \gtrsim 10^3$\,TeV were detected among the families, while two other spectra, both with lesser and larger $\overline{N_e}$, terminate already at an upper $E_\gamma$ limit of about $(100-150)$\,TeV. Moreover, even in the $\overline{N_e}=(5\cdot 10^6)$ range spectrum the last point which corresponds to the most energetic gamma ray quanta is separated from the preceding part of the distribution by a noticeable gap of zero bins which spans from $E_\gamma$  values of $(200-300)$\,TeV until $E_\gamma \sim 1000$\,TeV. Such \ch{a} considerably detached outlier might arise either as a result of systematic error \ch{in determination of electromagnetic cascade energy by the photo-densitometry method of X-ray films}, or because of some non-trivial physical reason.

\ch{As for the possibility of densitometry fault}, it should be \ch{firstly} noted that the relation between \ch{the} $E_\gamma$ values of \ch{the} two \ch{trailing} points in the $\overline{N_e}=(5\cdot 10^6)$ distribution is about $5$. This is an order of magnitude above the typical inaccuracy of energy estimation by \ch{the} XREC method shown in Fig.~\ref{figishaulov_xrecerrs} \ch{for the TeV scale gamma rays}. Though so large miscalculation in energy estimate for \ch{the last point of the middle distribution} already seems unlikely, a more strict check can be made in the considered case. For the purpose, the indices $\beta$ were defined of \ch{the} power approximations $dN_\gamma/dN_\gamma \sim E^{-\beta}$ drawn over the falling parts of each spectrum in Fig.~\ref{figishaulov_spcd}. For better reliability, this calculation was made twice: as usual linear approximation by the least mean square method (LMS) of the experimental points plotted in double logarithmic scale, and by minimization of a $\chi^2$ type sum of \ch{the} discrepancies between the mean energy values in each $E_\gamma$ bin and approximating power function. In both variants the very last outlier point of the $\overline{N_e}=(5\cdot 10^6)$ distribution did not participate in the approximation procedure.

\begin{table}
\begin{center}
\caption{Best fit values for the slope index $\beta$ in power approximation $dN_\gamma/d E_\gamma \sim E_\gamma^{-\beta}$ of the falling parts of differential gamma ray spectra from Fig.~\ref{figishaulov_spcd}.}
\label{tabishaulov_spcd}
\begin{tabular*}{\columnwidth}{@{\extracolsep{\fill}}c|ccc}
\hline
 $N_e$ & $10^6$ & $5\cdot 10^6$ & $6\cdot 10^7$
\\
\hline

% #2
L.M.S. & $2.2\pm 0.2$ & $1.2\pm 0.1$ & $1.7\pm 0.2$
\\

% #1
$\chi^2$ & $1.9\pm 0.1$ & $1.26\pm 0.09$ & $1.7\pm 0.2$
\\

\hline
\end{tabular*}
\end{center}
\end{table}

The results of both calculation types are listed in Table~\ref{tabishaulov_spcd}, and in Fig.~\ref{figishaulov_spcd} they are drawn by the straight lines: the solid lines correspond to the LMS type estimations, while the dotted ones represent the fitting of a $\chi^2$ functional. As it is seen there, the difference between the two approximation methods is comparable with their own errors, so it can be said that both produce the same result. As well, the last point of the $\overline{N_e}=(5\cdot 10^6)$ spectrum \ch{is} indeed compatible with the common power law, since the approximation lines \ch{intersect} its error bars. This is the reason to consider this point as trustful.

\begin{figure*}
{\centering
\includegraphics[width=0.6\textwidth, trim=0mm 0mm 0mm 0mm]{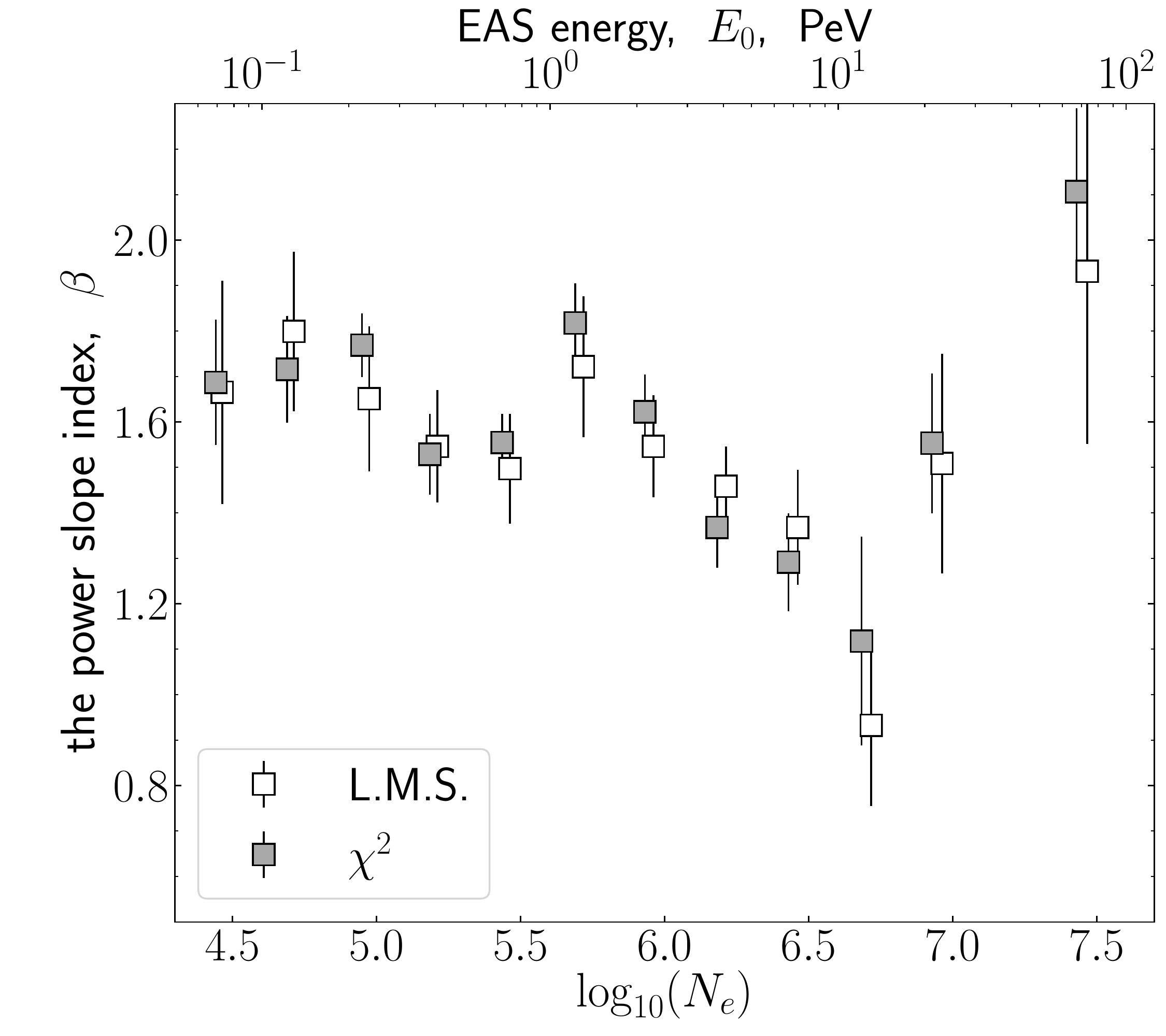}
\caption{The dependence of the power approximation index $\beta$ of the falling part of differential gamma ray spectrum in families on the mean size $N_e$ of accompanying shower. The upper axis is graduated in accordance with the $E_0=(2.5\cdot10^{-6}\times N_e)$\,PeV re-calculation rule.
}
\label{figishaulov_spcdb}}
\end{figure*}

Finding such peculiar prolongation of the $d N_\gamma/d E_\gamma$ distribution \ch{which existed only} in a limited range of $N_e$ values was a motif for more precise investigation of the shape of gamma ray spectra in \ch{the} families. For this purpose, the whole statistics of family events was splitted into a number of groups correspondingly to the size of accompanying EAS, and the indices $\beta$ were defined for the falling part of \ch{the average} gamma ray \ch{spectrum} of  the events \ch{of every group}. As before, approximation of the experimental points by a power type function $dN_\gamma/dN_\gamma \sim E^{-\beta}$ was made by the two methods, LMS and fitting of a $\chi^2$ sum. The result of this operation is presented in Fig.~\ref{figishaulov_spcdb} as dependence of $\beta$ indices in gamma families \ch{on the average size $N_e$ of corresponding EAS}.

According to Fig.~\ref{figishaulov_spcdb}, in the range of EAS sizes $N_e\simeq (10^6-10^7)$, \ch{\idest between the energy limits of $E_0\simeq 3$\,PeV and $E_0\simeq (20-30)$\,PeV}, the slope of \ch{the} differential energy spectra of gamma ray quanta in families does systematically diminish itself, but further on it returns to its initial, or somewhat larger, value \ch{among the events with $N_e\gtrsim 10^7$}. Though insufficient statistics of the events with $N_e\sim 10^8$ prevents making more decisive conclusion there, it can be said just now that such variation of spectrum shape is equivalent to an increase of the average energy of hadrons which constitute an EAS core, and this change takes place locally in a limited range of shower sizes of $N_e\simeq (10^6-10^7)$, or, in other words, among the EAS with primary energy $E_0$ between $3$\,PeV and $20$\,PeV.

\subsection{\ch{Energy distribution of the high energy hadrons of EAS core in comparison with simulated families}}

For \ch{further} analysis of the behavior of gamma ray quanta in families it is convenient to normalize their individual energies $E_\gamma$ to the primary energy $E_0$ of accompanying extensive air shower. Such procedure permits to avoid overlapping of the spectra drawn for different $E_0$ ranges on a same plot. \ch{Thus,} analogously to the Feynman variable $x_F$ mentioned in Introduction, in the case of XREC data it can be introduced a dimensionless variable $x_\gamma=E_\gamma/E_0$. As it was explained \ch{in Section\,\ref{sectieas}}, the estimations of EAS energy $E_0$ for such analysis can be calculated starting from the experimentally measured shower sizes $N_e$, as $E_0=(2.5\cdot10^{-6}\times N_e)$\,PeV \cite{ontien_icrc1987___through_ne_ru, tamada_simul_1994_e0_from_ne}.

Since the minimum energy of gamma ray quanta detected in families \ch{depends} only on the registration threshold of X-ray films and is the same for all $E_0$, the lower abscissa limit of \ch{the gamma ray spectra built over the $x_\gamma$ variable} must be in \ch{inverse} proportion to $E_0$ (and $N_e$). By variation of EAS energy \ch{such distributions} move in horizontal direction \ch{along the $x_\gamma$ axis}, and any deviation \ch{of their behavior from universal scaling type} can be quickly pinpointed as a change of \ch{the spectrum} slope or non-proportional displacement \ch{in horizontal direction}.

%%%%%%%%%%%%%%%%%%%%%%%%%%%%%%%%%%%
\ch{By investigation of the energy dependence of gamma ray spectrum} it is useful to exclude from consideration \ch{its} initial part which is distorted by the effect of registration threshold. As it was seen in Fig.~\ref{figishaulov_spcd}, the regular power shape \ch{in differential energy spectra} restores itself at an energy $E_\gamma\simeq 6$\,TeV, so by further analysis only the gamma ray quanta with energy $E_\gamma\geqslant 6$\,TeV were included into the spectra over \ch{the} $x_\gamma$ variable. 
%%%%%%%%%%%%%%%%%%%%%%%%%%%%%%%%%%%

\begin{figure*}
{\centering
\includegraphics[width=0.49\textwidth, trim=0mm 0mm -1mm 0mm]{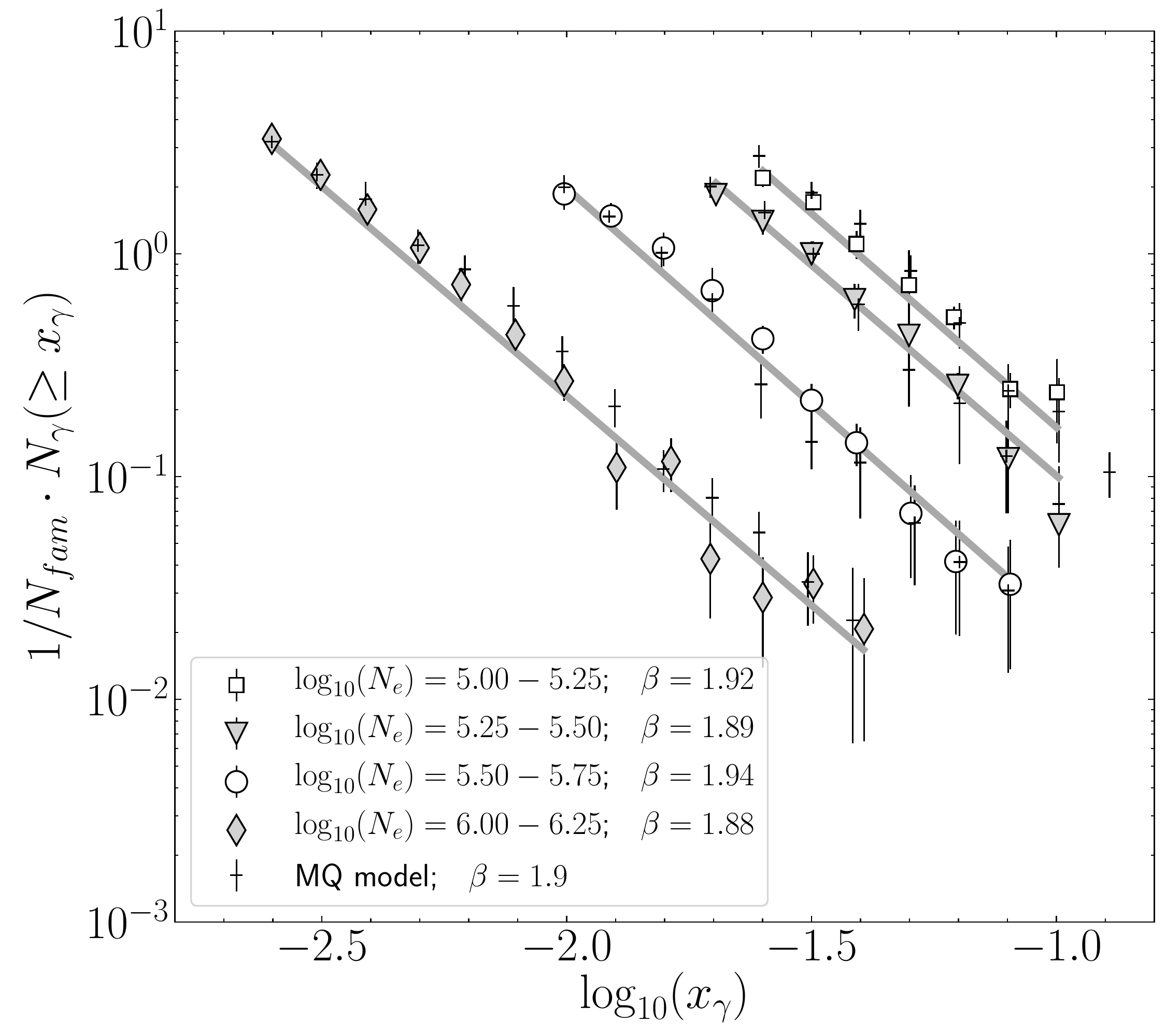}
\includegraphics[width=0.49\textwidth, trim=-1mm 0mm 0mm 0mm]{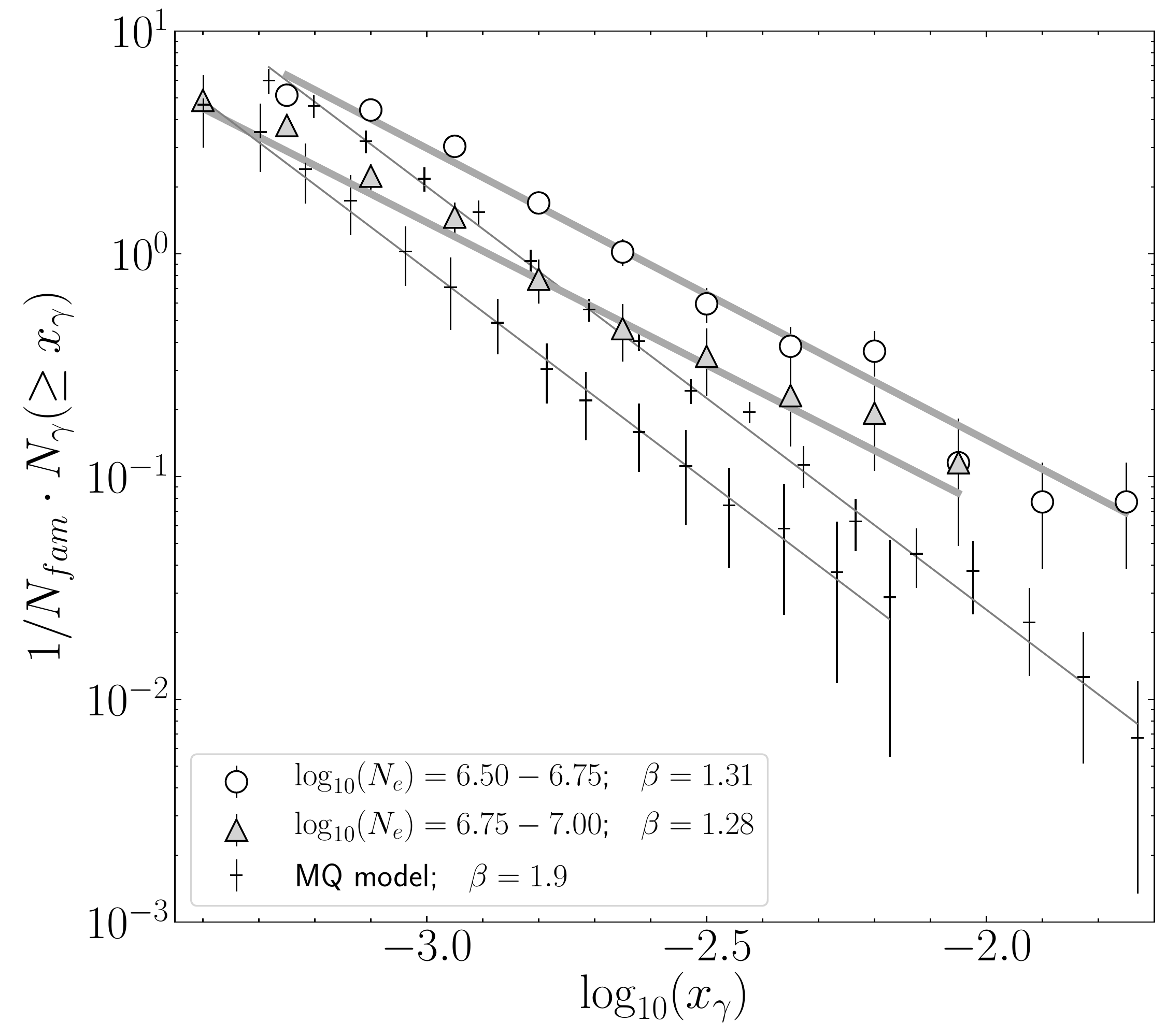}
\caption{\ch{Integral spectra, over dimensionless variable $x_\gamma=E_\gamma/E_0$, of gamma ray quanta in the families which belong to several log-equal intervals on the size $N_e$ of accompanying EAS. All spectra are normalized to the amount of gamma families $N_{fam}$ in corresponding $N_e$ interval. Straight lines indicate power approximation $N_\gamma(\geqslant $$x_\gamma) \sim x_\gamma^{-\beta}$ of the points in each spectrum. The criteria of XREC events selection for the plot were  $E_\gamma\geqslant 6$\,TeV, $N_\gamma\geqslant 1$, $\sum E_\gamma\geqslant 16$\,TeV. The spectra of gamma rays in simulated events \cite{onxrec_simulation_tien_hadron_1994} are shown with crosses beside each experimental distribution.}}
\label{figishaulov_spci_a}}
\end{figure*}

\ch{
Several integral spectra $N_\gamma(\geqslant $$x_\gamma)$ of the gamma ray quanta in the families correlated with corresponding showers is presented in Fig.~\ref{figishaulov_spci_a}. The spectra are averaged between the XREC--EAS events pairs which belong to the different EAS size intervals with a constant logarithmic step $\Delta \log_{10}(N_e)=0.25$. With such stepping the intervals occur certainly wider than the typical error of EAS size determination in the experiment.
Among the XREC events only the gamma ray quanta with the energy $E_\gamma\geqslant 6$\,TeV were considered by building these distributions. To be included into the spectra of the plot a family must contain at least one gamma ray quantum with sufficient energy ($N_\gamma\geqslant 1$), and the sum energy of all its quanta $\sum E_\gamma$ must be above $16$\,TeV. As before, the energy distributions in Fig.~\ref{figishaulov_spci_a}  were normalized to the number of gamma families $N_{fam}$ in the corresponding $N_e$ interval.
}

\ch{
Together with the measurement points, there are simulated spectra plotted with crosses in Fig.~\ref{figishaulov_spci_a} beside each experimental distribution. These are the results of simulation \cite{onxrec_simulation_tien_hadron_1994} which was made for the  process of family production in the Tien Shan XREC on the basis of the MQ~model of the high-energy particles interaction. The latter} \ch{was formerly elaborated by the \PAMIR collaboration for a mountain experiment of the same name and calibrated by its results \cite{onxrec_simulation_mq_1991}. In comparison \cite{onxrec_simulation_mq_qgs_1999} of the MQ model with a more contemporary combination QGS\-JET+COR\-SI\-KA it was stated mutual similarity between their predictions concerning the properties of resulting gamma families, as a consequence of the fact that in the ground of both simulation approaches lays the same quark-gluon strings mechanism of hadronic interaction. An important advantage of the MQ model is a built-in reckoning of the development of electromagnetic cascades and other technical features of the gamma ray registration process in XREC.
}

\ch{
As it is seen in the left plot frame of Fig.~\ref{figishaulov_spci_a}, up to the shower size value $N_e\simeq 10^6$ ($E_0\simeq 3$\,PeV) the energy spectra of gamma rays in experimental families can be characterized by scaling behavior which reveals through regular displacement of their mutually parallel plots in proportion with the energy growth. The slope index $\beta$ of a power approximation $N_\gamma(\geqslant x_\gamma)\sim x_\gamma^{-\beta}$ of the experimental data points in these leading $N_e$ intervals varies between the tight limits $(1.8-1.9)$, in good agreement with the value $\beta=(1.9\pm 0.1)$ predicted by the MQ simulation.  Close power index of the spectrum of secondary gamma rays was also reported for a somewhat lower energy range of $E_0\simeq(0.01-0.1)$\,PeV by a balloon experiment \cite{tev_gamma_spectra_1968} which used a hybrid detector consisting of the X-ray films and nuclear emulsion plates for investigation of hadronic interactions in the upper atmosphere.
}

\ch{
The scaling behavior breaks noticeably around the shower size value $N_e\simeq (3-6)\cdot 10^6$ (see the right panel plot in Fig.~\ref{figishaulov_spci_a}). In the families accompanied by the EAS which belong to this $N_e$ range the spectra of gamma rays become essentially harder: the $\beta$ index of} \ch{their power approximation drops down to $\sim$$1.3$. At the same time, the model spectra simulated for the EAS with corresponding primary energies $E_0\simeq (3-15)$\,PeV remain having the same slope around $\beta\simeq 1.9$ as before.
}

\begin{figure*}
{\centering
\includegraphics[width=\textwidth, trim=0mm 0mm 0mm -3mm]{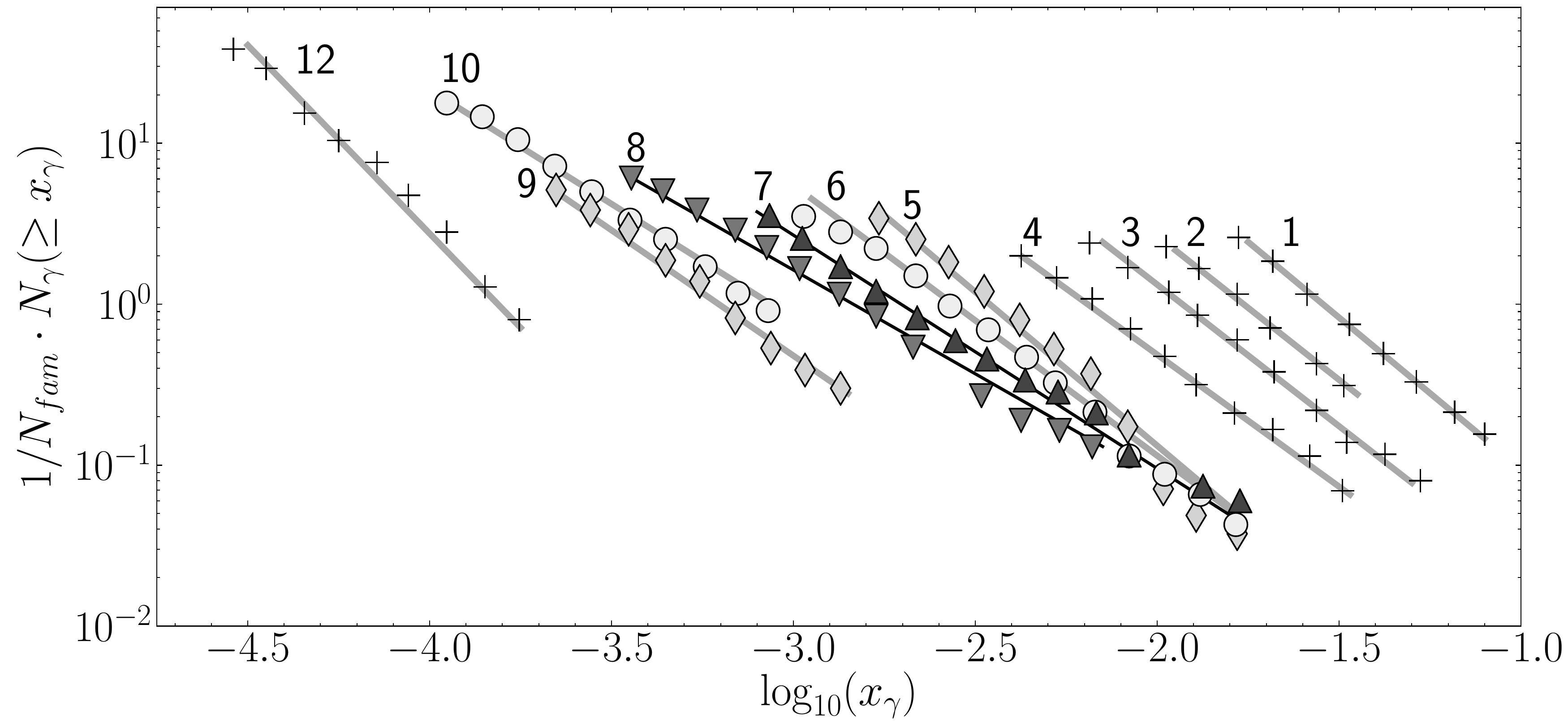}
\caption{\ch{Same distributions as in Fig.~\ref{figishaulov_spci_a} built for a wider variation range of shower sizes.} The digits beside \ch{each distribution} mean \ch{the ordering} numbers of the $N_e$ intervals listed in Table~\ref{tabishaulov_spci}. Straight lines \ch{correspond to} the power approximation of experimental spectra. \ch{To avoid confusion between curves, the error bars of experimental points are not shown.}}
\label{figishaulov_spci}}
\end{figure*}

\begin{table}
\begin{center}
\caption{
The limits of EAS size intervals, per-interval statistics of the families ($N_{fam}$) and gamma ray quanta ($N_{\gamma}$) for the threshold conditions $E_\gamma\geqslant 6$\,TeV, \ch{$\sum E_\gamma\geqslant 10$\,TeV}, and the slope indices $\beta$ in power approximation $N_\gamma(\geqslant $$x_\gamma) \sim x_\gamma^{-\beta}$ of the integral spectra from Fig.~\ref{figishaulov_spci}.
}
\label{tabishaulov_spci}
%\begin{tabular*}{\columnwidth}{@{\extracolsep{\fill}}cccccc}
\begin{tabular*}{\columnwidth}{@{\extracolsep{\fill}}ccccc}
\hline
\# & $\log_{10}(N_e)$ & $N_{fam}$ & $N_\gamma$ & $\beta$ %& $\beta_{\chi^2}$
\\
\hline

$1$ &
$4.93-5.18$ &
$119$ &
$175$ &
$1.88\pm 0.12$% &$1.79\pm 0.13$
\\

$2$ &
$5.18-5.43$ &
$164$ &
$248$ &
$1.79\pm 0.12$ %&$1.71\pm 0.10$
\\

$3$ &
$5.43-5.68$ &
$181$ &
$300$ &
$1.76\pm 0.08$ %&$1.64\pm 0.09$
\\

$4$ &
$5.68-5.93$ &
$176$ &
$314$ &
$1.64\pm 0.08$ %&$1.57\pm 0.08$
\\

$5$ &
$5.93-6.18$ &
$154$ &
$290$ &
$1.91\pm 0.11$ %&$1.86\pm 0.07$
\\

$6$ &
$6.18-6.43$ &
$149$ &
$393$ &
$1.68\pm 0.07$ %&$1.57\pm 0.10$
\\

$7$ &
$6.43-6.68$ &
$91$ &
$315$ &
$1.45\pm 0.06$ %&$1.50\pm 0.12$
\\

$8$  &
$6.68-6.93$ &
$46$ &
$194$ &
$1.19\pm 0.08$ %&$1.35\pm 0.13$
\\

$9$ &
$6.93-7.18$ &
$16$ &
$62$ &
$1.42\pm 0.12$ %&$1.23\pm 0.18$
\\

$10$&
$7.18-7.43$ &
$11$ &
$89$ &
$1.56\pm 0.15$ %&$1.49\pm 0.27$
\\

$11$ &
$7.43-7.68$ &
$0$ &
$0$ &
--- %&---
\\

$12$ &
$7.68-7.93$ &
$4$ &
$56$ &
$2.34\pm 0.33$ %&$2.22\pm 0.10$
\\

\hline
\end{tabular*}
\end{center}
\end{table}

\ch{
A larger set of the gamma ray energy spectra in families is presented in the plot of} \ch{Fig.~\ref{figishaulov_spci}, where the experimental data points are drawn without their error bars for better distinctiveness of the neighbouring curves. Also, somewhat different borders were used in this figure at splitting the whole statistics into $12$ log-equal intervals on the parameter $N_e$ to stress in the best way the dependency of the spectrum shape on shower size.} A list \ch{of these borders} is given in Table~\ref{tabishaulov_spci}, together with the total count of the family events and gamma ray quanta \ch{in each $N_e$ interval}, and with estimations of the power \ch{spectrum index} $\beta$ which were defined \ch{by LMS} approximation of the experimental points in Fig.~\ref{figishaulov_spci}.

\ch{As before,} it is seen in Fig.~\ref{figishaulov_spci} \ch{that} over the four leading  $N_e$ intervals the gamma ray spectra keep scaling behavior. The scaling breaks in the intervals \#$(5-6)$, \idest at $N_e\simeq (1-3)\cdot 10^6$ and $E_0\simeq (3-6)$\,PeV. In the families which belong to the 7th and 8th interval, between $N_e\simeq (3-10)\cdot 10^6$ and $E_0\simeq (6-20)$\,PeV, the gamma ray spectra become essentially harder, with noticeable decrease of curves inclination in Fig.~\ref{figishaulov_spci} and diminishing of $\beta$ indices in Table~\ref{tabishaulov_spci}. Among the most powerful EAS of 12th interval, those with $N_e\geqslant 5\cdot 10^7$  ($E_0\geqslant 100$\,PeV), the \ch{$x_\gamma$ spectrum} restores \ch{the} initial soft shape with $\beta \gtrsim (1.8-1.9)$, \ch{or becomes even softer}.

\ch{Since the integral gamma ray spectra $N_\gamma(\geqslant x_\gamma)$} in Fig.~\ref{figishaulov_spci} were normalized to the \ch{number of families $N_{fam}$ in each $N_e$ interval, the ordinate of its leftmost upper point equals precisely to the mean multiplicity of gamma ray quanta $\overline{N_\gamma}=\sum N_\gamma/N_{fam}$ in the families of that interval}. Disproportional displacement of $10$th spectrum along the vertical axis \ch{in this figure} indicates a sudden local rise of the mean multiplicity of high-energy gamma rays \ch{in the EAS with $N_e\simeq (1.5-3)\cdot 10^7$, \idest at $E_0\simeq (40-70)$\,PeV}. This is another sign of non-uniformity \ch{in the energy dependence of the production process of gamma rays (and neutral pions) at interaction of the PeV scale cosmic ray particles}.

In sum, the data of the \HADRON experiment did show that the spectra of the high energy gamma ray quanta in families, and consequently of the most energetic hadrons of EAS core, can not be represented in scaling manner by a single universal function which would keep a same slope over the whole range of investigated primary energies, $E_0\simeq (0.2-200)$\,PeV. Actually, the power slope index of gamma ray spectra varies non-monotonously with increase of the average size of EAS: up to $N_e\simeq 10^6$ it keeps, indeed, steady around the value $\beta\simeq (1.7-1.9)$ \ch{which is compatible with a simulation based on a scaling type MQ model}, then it starts to fall down and achieves its minimum of $\beta\simeq 1.2$ in the showers with $N_e\simeq (5-8)\cdot 10^6$, where the gamma rays spectra acquire a most hard shape, then it returns to its initial or somewhat larger value, and the spectra become softer again. 

\begin{figure*}
{\centering
\includegraphics[width=0.57\textwidth, trim=0mm 0mm 0mm 0mm]{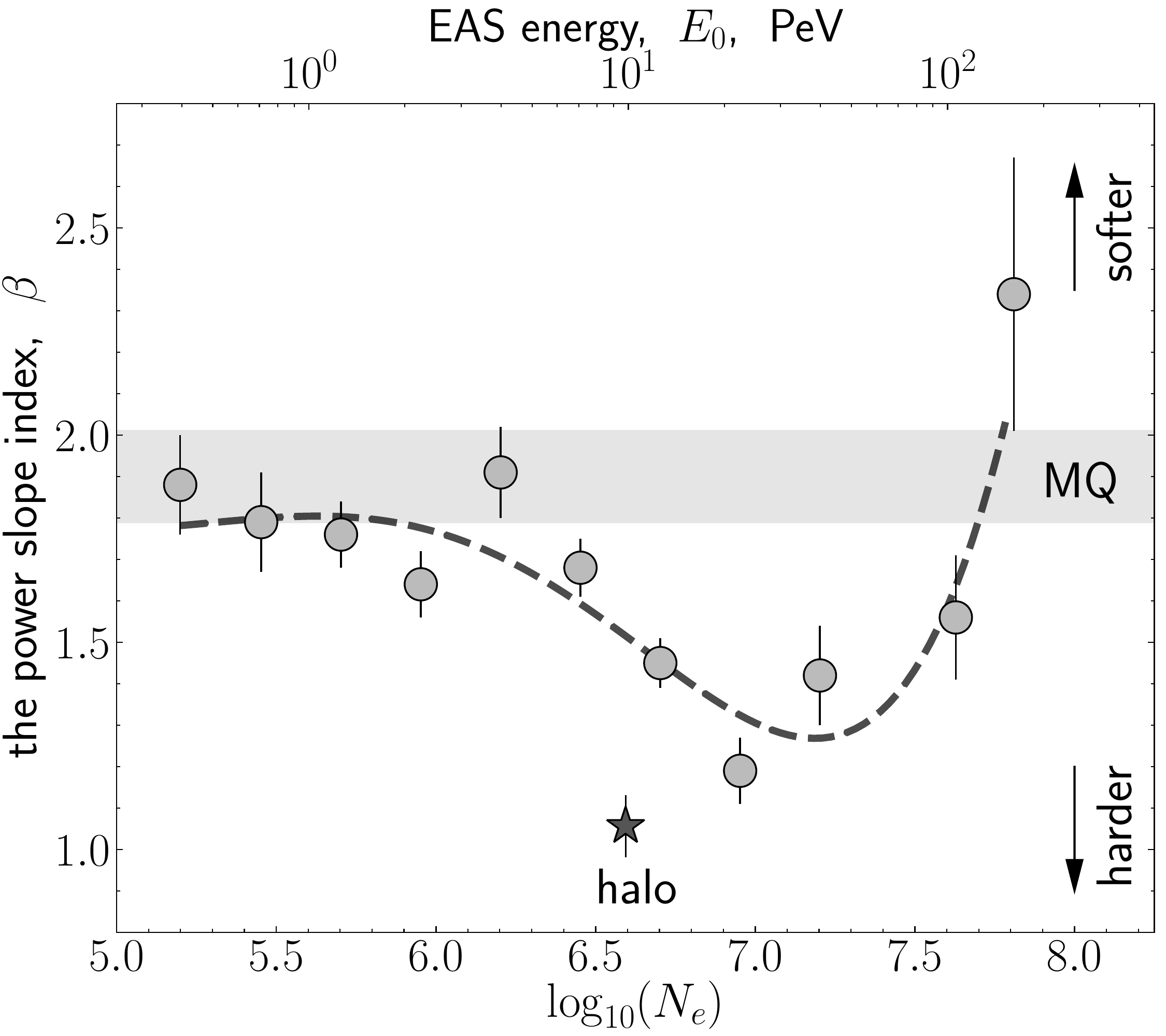}
\caption{Power indices $\beta$ of the approximation lines of integral spectra $N_\gamma(\geqslant $$x_\gamma)$ in Fig.~\ref{figishaulov_spci}, in dependence on the EAS size variable $N_e$. The wide horizontal strip corresponds to prediction of the MQ simulation model. \ch{The smooth dashed curve represents a polynomial fit of experimental points: $0.00269x^6 - 0.0593x^5 + 0.459x^4 - 1.31x^3 + 0.0258x^2 + 3.58x + 3.57$, where $x=\log_{10}N_e$. The mean $N_e$ and $\beta$ values in the families with halo (see Section\,\ref{sectihalo}) are indicated with a star.}}
\label{figishaulov_spcib}}
\end{figure*}

\ch{
The variation of the power slope index of the gamma ray spectra  in dependence on the size of corresponding shower is reproduced by the plot in Fig.~\ref{figishaulov_spcib}. The continuous dashed line in that figure represents a smoothed polynomial fit of the experimental points, while the simulation result \cite{onxrec_simulation_tien_hadron_1994} is shown there by a wide horizontal strip. According to the MQ model, the slope index of the integral gamma ray spectrum in families must keep a constant value $\beta=(1.9\pm 0.1)$ in the whole variation range of EAS energy  between $1$\,PeV and $200$\,PeV. In contrary, as it follows from the figure, the experimental points do agree with the simulation only in the showers with $N_e\lesssim 3\cdot 10^6$, \idest until the energy} \ch{of $E_0\simeq (5-7)$\,PeV, and then the scaling behavior of the gamma ray spectra obviously terminates.
}

\ch{Any} deviation of the experimental data points from \ch{the} universal power law predicted by simulation could be imitated erroneously, \ch{either} because of \ch{some} error \ch{in} determination of the gamma ray spectra slope in experiment, or \ch{by} statistical fluctuations of the Monte-Carlo results. The probability of such trivial \ch{explanation of the scattering of experimental points in Fig.~\ref{figishaulov_spcib}} can be estimated in usual way, by calculation of \ch{a} discrepancy sum
$$
\chi^2=\sum_i{\left(\frac{\beta_i-\beta_T}{\sigma}\right)^2},
$$
and its comparison with the probability $P$ of the standard $\chi^2$ distribution:
$$
P_n(\chi^2\geqslant \chi^2_0)=\frac{2}{2^{n/2}\Gamma(n/2)}\cdot \int\limits_{\chi_0}^{\infty} x^{n-1}e^{-x^2/2} dx.
$$
In these formulae $\beta_i$ are the experimental estimations of power slope in $n$ points of Fig.~\ref{figishaulov_spcib} ($i=1\ldots n$), and $\beta_T=1.9$ is the theoretical value which follows from the MQ simulation. As dispersion $\sigma$ in each point it can be used either the errors $\sigma_i$ of \ch{the slope estimates resulting from the experiment, as they were listed in} Table~\ref{tabishaulov_spci}, or the sum error of the experimental \ch{and simulated slope estimations}, $\sigma_{iT}=\sqrt{\sigma_i^2 + \sigma_T^2}$, where $\sigma_T=0.1$. 

\begin{table*}
\begin{center}
\caption{The $\chi^2$ discrepancy sums and corresponding probabilities $P$ of purely random deviation of the experimental points  in Fig.~\ref{figishaulov_spcib}  from a constant $\beta_T$ value as defined in various calculation variants (see text).
}
\label{tabichi}
\begin{tabular*}{\textwidth}{@{\extracolsep{\fill}}c|c|c|c|c|c}
\hline
\hline
$N_e$ &
$n$ &
\makecell{$\chi^2$, $P$\\for $\sigma_i^2$ only,\\$\beta_T=1.90$} &
\makecell{$\chi^2$, $P$,\\with $\sigma_{iT}^2$,\\$\beta_T=1.90$} &
\makecell{$\chi^2$, $P$\\for $\sigma_i^2$ only,\\$\beta_T=1.65$} &
\makecell{$\chi^2$, $P$,\\with $\sigma_{iT}^2$\\$\beta_T=1.65$}
\\
\hline
\hline
$10^5-10^8$&
$11$&
\makecell{$159$,\\$P_{11}\sim 10^{-28}$}&
\makecell{$64$,\\$P_{11}\simeq 2\cdot 10^{-9}$}&
\makecell{$58$,\\$P_{11}\simeq 1\cdot 10^{-8}$}&
\makecell{$48$,\\$P_{11}\simeq 7\cdot 10^{-7}$}
\\
\hline
$10^6-10^8$&
$7$&
\makecell{$144$,\\$P_{7}\sim 10^{-27}$}&
\makecell{$58$,\\$P_{7}\simeq 4\cdot 10^{-10}$}&
\makecell{$46$,\\$P_{7}\simeq 6\cdot 10^{-8}$}&
\makecell{$32$,\\$P_{7}\simeq 2\cdot 10^{-5}$}
\\
\hline
\makecell{$10^6-10^8$,\\ without \\ the edge \\ points}&
$5$&
\makecell{$142$,\\$P_{5}\sim 10^{-29}$}&
\makecell{$56$,\\$P_{5}\simeq 7\cdot 10^{-11}$}&
\makecell{$16$,\\$P_{5}\simeq 4\cdot 10^{-3}$}&
\makecell{$11$,\\$P_{5}\simeq 2\cdot 10^{-2}$}
\\
\hline
\hline
\end{tabular*}
\end{center}
\end{table*}

\ch{The $\chi^2$ sum values which have been} obtained for both \ch{variants of the dispersion calculation rule} are listed in the third and fourth columns of Table~\ref{tabichi}. For each $\sigma$ case, the calculation was made thrice: over the whole range of investigated $N_e$ values (with $n=11$ experimental points), only for the points with significant deviation from the expected $\beta_T$ value ($7$~points with $N_e$ between $10^6$ and $10^8$), and for the same interval of \ch{the} large $\beta_i$ deviations between $N_e=10^6$ and $N_e=10^8$ with its two border points excluded from summation ($n=5$). As it follows from Table~\ref{tabichi}, in any of these calculation variants the resulting probability to find the observed difference of the experimental points from the constant $\beta_T=1.9$ \ch{only because of} combination of purely random fluctuations occurs to be negligible.

It might be supposed also that it is the model put in the basis of the simulation which was wrong. \ch{In this case} the compatibility \ch{can be verified of the slope indices $\beta$ in} the large-sized $N_e$ intervals of Fig.~\ref{figishaulov_spcib} with a single constant level $\beta_T=1.65$, which is an average calculated over all points between the $N_e$ limits $10^6$ and $10^8$.
% (                     1.91+1.68+1.45+1.42+1.19+1.56+2.34)/7
\ch{The corresponding values of} $\chi^2$ sum and probability $P$ are shown in the $5$th and $6$th columns of Table~\ref{tabichi}. As it follows from these estimations, the experimental points remain incompatible, again, with the constant because of insignificant probability, $P\lesssim (10^{-3}-10^{-2})$ in any case.

Hence, \ch{a supposition on possible imitation of the dispersion of experimental points in Fig.~\ref{figishaulov_spcib} due to random fluctuations does not sustain statistical criterion, and alternative conclusion on existence of some physically based  reason for} deviation from the universal scaling \ch{form} of the experimental gamma ray spectra occurs to be reliably founded. \ch{As a consequence, the same can be stated about the spectra of the high energy hadrons of EAS core which are predecessors of the detected gamma rays.}

\ch{
It should be noted that the lower border of the shower size region where disappearance of scaling behavior starts to be evident among the gamma ray spectra} \ch{in XREC families, $N_e\simeq 3\cdot 10^6$, is close to the position of the knee in the cosmic ray spectrum at $E_0\simeq 3$\,PeV, \idest $N_e\simeq 10^6$. Such a coincidence could hardly be random, and it is rather likely that both effects should have a same physical reason.
}

\subsection{The energy range of scaling violation}

The rise of the relative share of high-energy gamma ray quanta in  families, \ch{together} with corresponding drop of the slope of their spectrum, may be caused either by appearance of the events with an absence of low-energy gamma rays, or by general hardening of the gamma ray spectrum. An analysis of these possibilities supposes selection, accordingly to some formal rule, of a special group of events responsible for generation of the hard spectra. A proper criterion for the purpose seems to be the maximum energy of the gamma ray quanta in a family, $E_\gamma^{M}$, or the highest value $x_\gamma^{M}=E_\gamma^{M}/E_0$ in each particular spectrum (as before, $E_0$ is the energy of accompanying EAS). In EAS events the $x_\gamma^{M}$ parameter is \ch{evidently} determined by the dissipation \ch{rate} of the primary \ch{cosmic ray energy} in a nuclear-electromagnetic cascade which develops above the XREC installation. Since the gamma ray families stem from the secondary EAS hadrons produced \ch{in cascade} at a height of $(2-5)$\,km, both the shape of $x_\gamma$ distribution and its maximum value $x_\gamma^{M}$ depend on the properties of nuclear interaction and, in even larger degree, on the composition of primary cosmic ray particles.

\begin{figure*}
{\centering
\includegraphics[width=0.6\textwidth, trim=0mm 0mm 0mm 0mm]{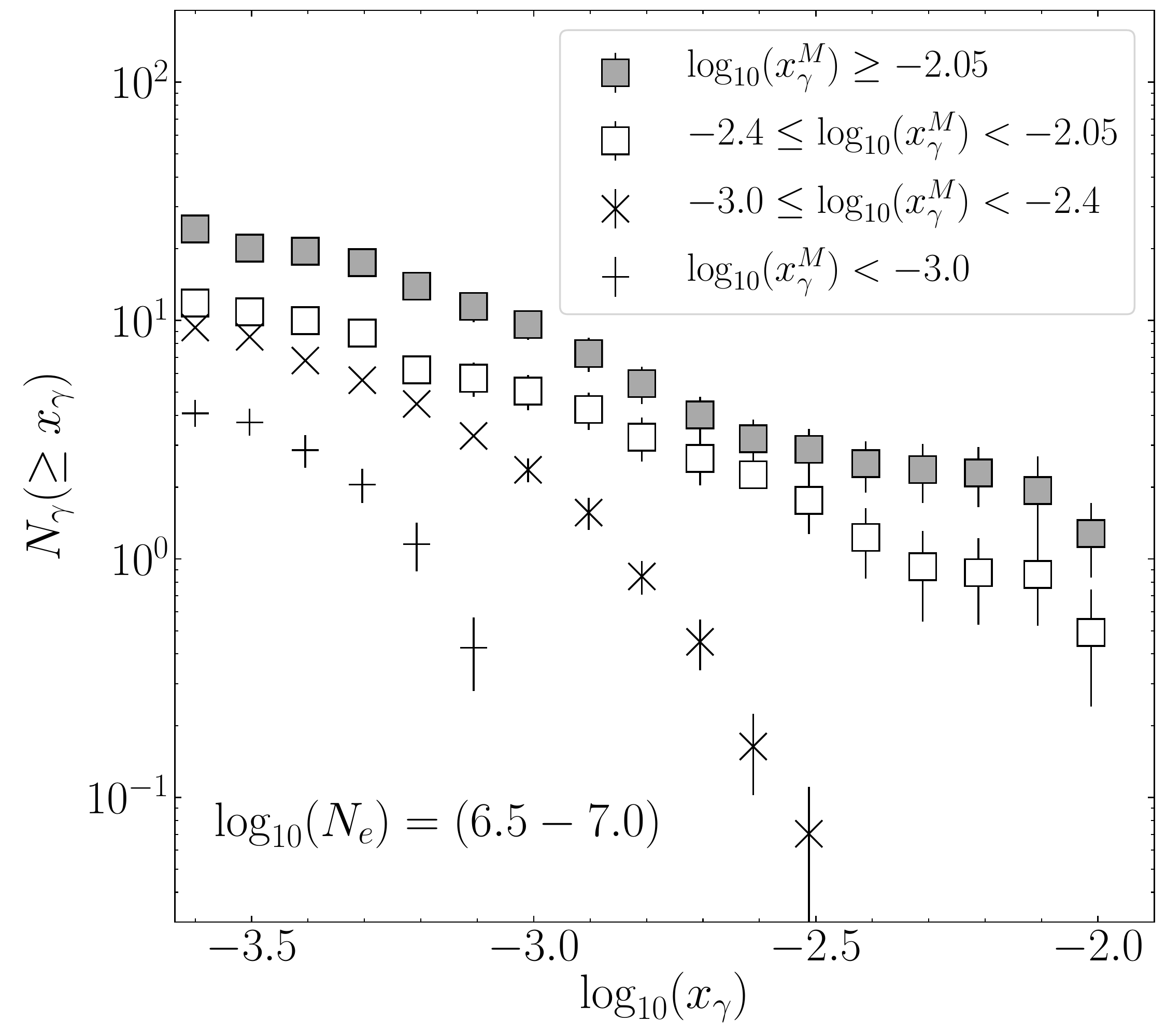}
\caption{Integral spectra \ch{of gamma rays in the families accompanied by the EAS} with $N_e\simeq (3\cdot 10^6-10^7)$ (\idest $E_0\simeq (8-25)$\,PeV) grouped by \ch{the maximum energy limit $x_\gamma^M$ of the detected gamma ray quanta}.}
\label{figishaulov_spcic}}
\end{figure*}

Practically all spectra shown in Fig.~\ref{figishaulov_spci} include \ch{some} gamma families with large values of $x_\gamma$ but relative proportion of such events generally is of a few percent only. It is the EAS size range between the limits of $N_e=3\cdot 10^6$ and $N_e=10^7$ (which mostly includes the $7$th and $8$th $N_e$ intervals of Table~\ref{tabishaulov_spci} and Fig.~\ref{figishaulov_spci}) where such families constitute a noticeable part among all detected events. The whole statistics of family events which belong to this range of shower sizes is sufficiently \ch{rich} to divide them between \ch{the} two groups with $\log_{10}(x_\gamma^{M})<-2.4$ and $\log_{10}(x_\gamma^{M})\geqslant -2.4$. The number of events in these groups is, correspondingly, $77$ and $18$.

Strictly speaking, any comparison of energy spectra calculated for the families of both mentioned groups might be not completely adequate since the first $x_\gamma^{M}$ interval is limited by the border of $\log_{10}(x_\gamma^{M})=-2.4$, while the other one is open from above. To check \ch{the} possible influence of the events selection criterion on the shape of resulting \ch{energy distribution} each of the $x_\gamma^{M}$ intervals can be additionally splitted into two parts. The integral spectra \ch{over the value $x_\gamma$}  built for the families of all four events groups are presented in the plot of Fig.~\ref{figishaulov_spcic}.

As it follows from Fig.~\ref{figishaulov_spcic}, in the families which belong to the two groups with $\log_{10}(x_\gamma^{M})<-2.4$ the spectra of gamma ray quanta are of an essentially non-power kind and terminate abruptly even before their limiting values $x_\gamma^{M}$. In contrary, in \ch{the} families with $\log_{10}(x_\gamma^{M})\geqslant -2.4$ the spectra have a much more regular shape which is close to power function with a single slope index over the whole range of $x_\gamma$ variation. This conclusion remains valid both for the events selection interval $-2.4\leqslant \log_{10}(x_\gamma^{M})< -2.05$ restricted from above, and for the open group of events with $\log_{10}(x_\gamma^{M})\geqslant -2.05$ which has not any fixed upper limit.

To sum up, among the EAS events which belong to one and the same relatively narrow diapason of shower sizes $N_e$ there were found the families both with a soft and a hard, power-like shape of the energy spectrum of particular gamma ray quanta. A simple formal criterion to distinguish between these two essentially different groups of events is comparison of the highest $x_\gamma^M$ value in gamma ray spectrum with the border limit of $10^{-2.4}$ (\idest $\simeq$$0.004$): all events with $x_\gamma^{M}$ parameter exceeding this limit relate to a specific group of families with extremely hard spectrum, which \ch{fact} is an evident sign of scaling violation. It is admixture of such cases to the whole statistics of gamma families \ch{that} is responsible for the hard gamma ray spectra of the $7$th and $8$th $N_e$ intervals in Fig.~\ref{figishaulov_spci}.

\begin{figure*}
{\centering
\includegraphics[width=0.48\textwidth, trim=0mm 0mm 0mm 0mm]{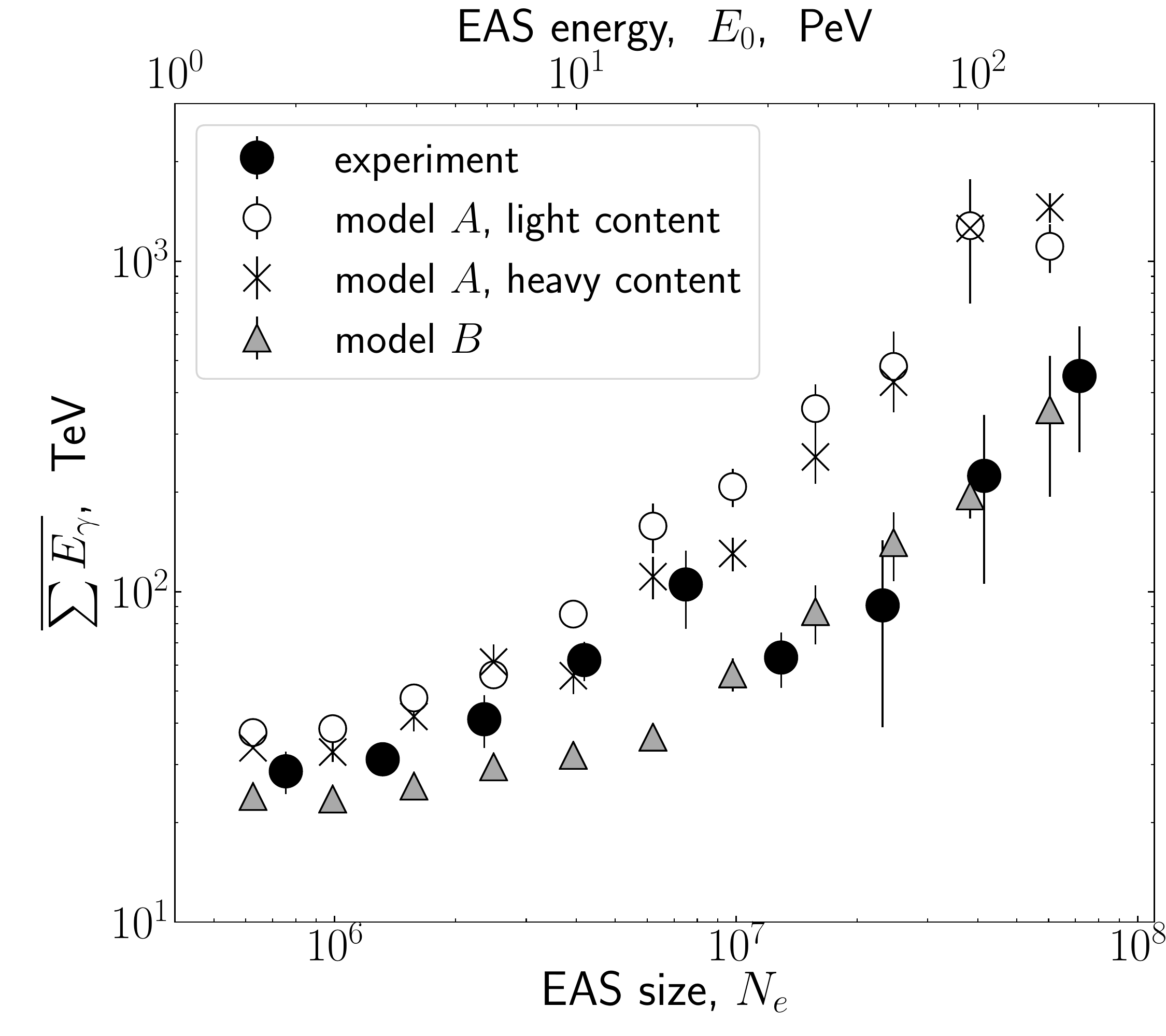}\,\,
\includegraphics[width=0.48\textwidth, trim=0mm 0mm 0mm 0mm]{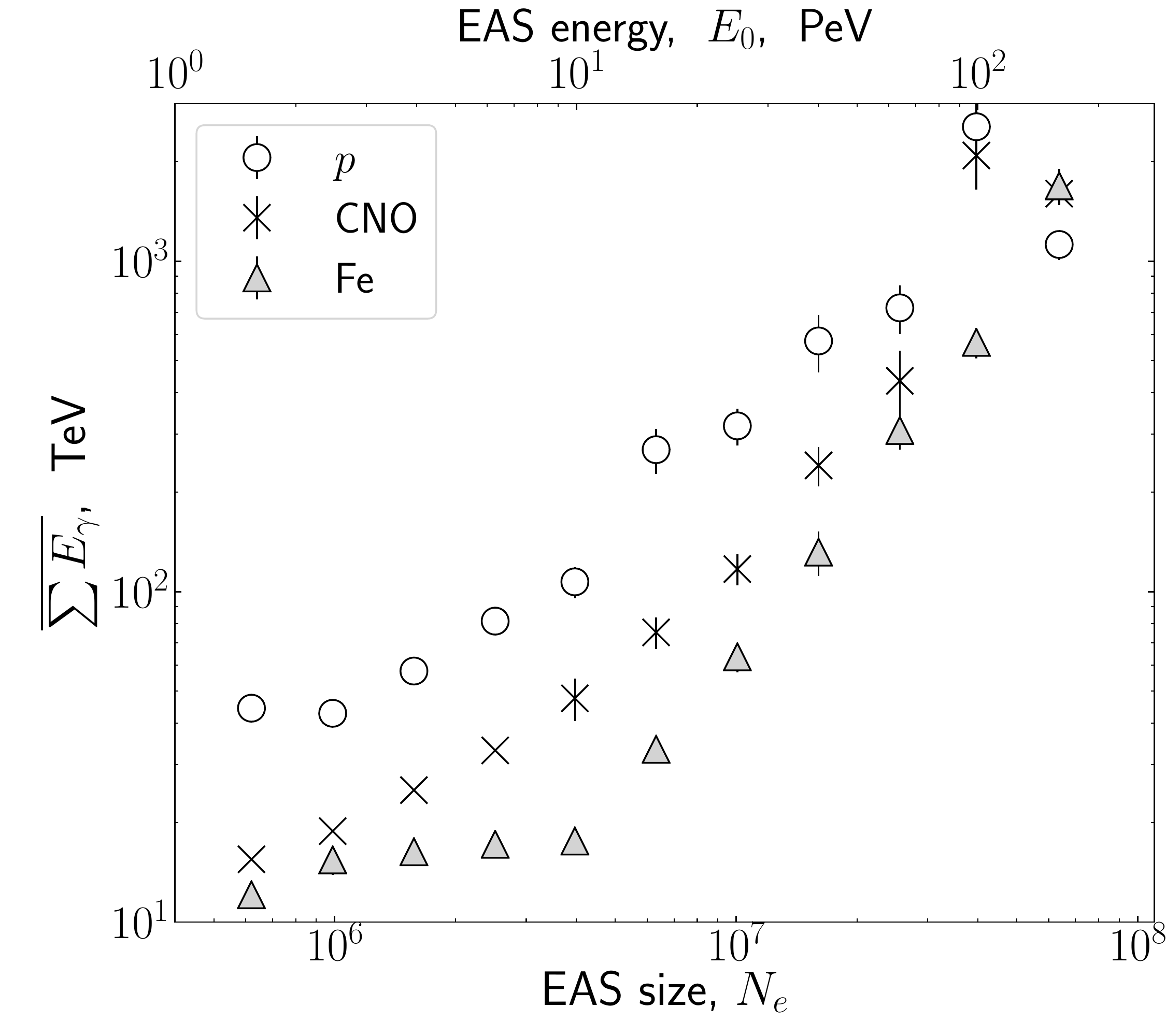}
\caption{Average sum energy of the gamma ray quanta in families, in dependence on the size of accompanying EAS. Left: comparison of experimental data points with the simulation \cite{tamada_simul_1994_e0_from_ne}. Right: the results of \ch{the} simulation \cite{tamada_simul_1994_e0_from_ne} for \ch{various} types of primary particles.}
\label{figitamada_nf_vs_ne}}
\end{figure*}

Since at the large $N_e$ values the statistics of EAS events with families is generally scarce, for precise localization of the region of scaling violation on the energy scale of cosmic rays it \ch{might} be more convenient to use the dependence of the average sum energy of \ch{all} gamma rays in families $\overline{\sum{E_\gamma}}$ on $N_e$, instead of the energy spectra of particular quanta. Such dependence is presented, with bold black circles, in the left plot of Fig.~\ref{figitamada_nf_vs_ne}. As it follows from this plot, monotonous increase of the sum gamma ray energy in families
with rising $N_e$ is sharply interrupted between the EAS size values $N_e\simeq(8-10)\cdot 10^{6}$, \ch{\idest at} $E_0\simeq (20-30)$\,PeV, where the $\overline{\sum{E_\gamma}}$ sum drops suddenly from $\sim$$100$\,TeV to $\sim$$50$\,TeV.

Experimental data on $\overline{\sum{E_\gamma}}$ in Fig.~\ref{figitamada_nf_vs_ne} are presented in comparison with corresponding results of \ch{the} simulation \cite{tamada_simul_1994_e0_from_ne}. The latter was made using an UA5 model of particles interaction (model $A$) which was based on extrapolation of contemporary accelerator data, and for \ch{the} two suppositions concerning the chemical composition of primary cosmic rays: a ``light'' content, with $\sim$$60$\% share of protons and alpha-particles, and a heavy one, with a $\sim$$(40-50)$\% predominance of Fe nuclei. Another model ($B$) anticipated forced production of the exotic ``Centauro species'' events \cite{centauro0,centauro1,centauro2}, with the relative share of energy transfer into secondary gamma ray quanta reduced to an order of magnitude. It follows from Fig.~\ref{figitamada_nf_vs_ne} that neither \ch{of two models} does well describe the experiment in the whole range of investigated shower sizes: up to $N_e\simeq 8\cdot 10^{6}$ ($E_0\simeq 20$\,PeV) the experimental data points remain close to prediction of \ch{the} model $A$, further on they comply mostly with \ch{the} model $B$. This result agrees with the above supposition about existence of two components with different interaction properties in the common flux of cosmic rays. According to disposition of the experimental points in the left hand plot of Fig.~\ref{figitamada_nf_vs_ne}, the region of obvious scaling violation due to \ch{the} presence of \ch{an} additional component in cosmic rays must be limited between the borders of $N_e\simeq (1-8)\cdot 10^6$, or $E_0\simeq (3-20)$\,PeV.

\subsection{The gamma ray spectrum in the families with halo}
\label{sectihalo}

Some gamma families detected in the \HADRON experiment have diffuse continuous darkening spot in their center, \ch{commonly named \textit{halo}} \cite{halo1,onxrec_mass_2017,onxrec_mass_2018}. The \ch{peak} of the EAS size distribution among the events with halo falls just into the region of scaling violation between $N_e\simeq(3-10)\cdot 10^6$, and the average slope of gamma ray spectrum \ch{in these events} is $\beta=(1.03\pm 0.07)$; in the coordinate frame of Fig.~\ref{figishaulov_spcib} both values are designated by a star marker. As a rule, halos appear in \ch{the} families with the highest sum energy of gamma ray quanta, $\sum{E_\gamma^{halo}} \gtrsim 100$\,TeV, which is essentially above the average $\overline{\sum{E_\gamma }}\simeq 40$\,TeV \ch{typical for the EAS} with the same primary energy $E_0$. As well, the average multiplicity of gamma ray quanta in \ch{the} families with halo is $\sim$$4$~times as large as in general (correspondingly, $(24.8\pm 4.1)$ \ch{against} $(6.6\pm 2.1)$). Because of large total energy, the whole statistics of \ch{the families with} halo is rather scarce, and amounts to 25~events only.

Generally, any local hardening of gamma ray spectrum (\idest the drop of its $\beta$ index in Fig.~\ref{figishaulov_spcib}) among \ch{a particular group of gamma families} is equivalent to increase of the average hadron energy $\overline{E_h}$ \ch{in the EAS connected with these events}. It is this effect that \ch{manifests} itself through deviation from scaling behavior in the \ch{gamma ray spectra of the families which correspond to the EAS with the sizes} between $N_e\simeq(1-10)\cdot 10^6$, or with the primary energy $E_0\simeq (3-20)$\,PeV. An appearance in the same energy range of the events with anomalously high $\sum{E_\gamma}$ (and consequently $\overline{E_h}$), and \ch{with} peculiarly large density of particles \ch{in} their cores, actually sufficient for halo production, is another evidence of the presence in the region of scaling violation of some specific class of EAS events.

\section{Discussion}

Joining an X-ray emulsion chamber and a shower installation in a single hybrid experimental complex \HADRON has for the first time permitted to measure directly the dependence of the spectra of secondary hadrons on the interaction energy $E_0$ of PeV order. It was found there significant change of the power spectrum slope of  secondary particles, \idest deviation from \ch{the} universal scaling-compatible spectrum shape, which takes place only within a limited range of $E_0$ values, $E_0\simeq (3-20)$\,PeV.

On the other hand, the results of analogous investigation of the secondary particles spectra currently available from the modern accelerator experiments, such as LHCf, do agree, actually, with anticipated scaling behavior of these spectra in the same PeV energy range. This new experimental fact means that the evident effect of scaling violation detected in the cosmic ray experiment can not be caused by any modification of the intrinsic features of particles interaction process, and its reason must be searched for among \ch{ composition of primary} cosmic rays.

%In the frames of possible modification of nuclear content the observed hardening of gamma ray spectra in the families can be explained only by increase of the relative share of light particles, such as protons and helium nuclei. According to the experiment, this effect must reveal itself starting from the primary energy of about $E_0\simeq 3$\,PeV, and below this  threshold the cosmic ray content must remain to be relatively heavy.

%%%% WHICH SIMULATIONS?
%The simulations show that the average slope of secondary hadron spectra $\beta$ generally depends on the type of primary particle. Thus, for iron primaries $\beta\simeq 2$, and for protons it is close to $1$.
Since the complex cosmic ray nuclei disintegrate to separate nucleons when coming into \ch{the} atmosphere, the spectrum slope of secondary hadrons generally depends on the type of cosmic ray particle: for a heavy primary nucleus the spectrum would be softer than for proton at the same interaction energy.
Experimental values of $\beta\simeq 1.9$ before the $3$\,PeV border and $\beta\simeq 1.2$ afterwards (see Table~\ref{tabishaulov_spci}) seem to be more likely consistent with a heavy content (Fe) below $3$\,PeV and practically pure mixture of \ch{protons} and He \ch{nuclei} at $E_0\simeq 20$\,PeV. Such a composition model could be assumed only in the case if the magnetic cut-off rigidity \ch{$R_{c}$ by acceleration of} individual nuclei were of about $0.1$\,PV, so that at $E_0\simeq 3$\,PeV cosmic rays generally consist of iron. Then, the observed scaling violation in gamma families can be explained by threshold appearance of an additional proton component above $3$\,PeV.

Some proof of this interpretation can be deduced from the dependence of the sum gamma ray energy in families $\sum{E_\gamma}$ on the shower size $N_e$ (\idest on the primary energy $E_0$) in Fig.~\ref{figitamada_nf_vs_ne}. The conclusion made in \cite{tamada_simul_1994_e0_from_ne} that neither of models $A$ and $B$ corresponds to experiment does not take into account the effect of scaling violation in actual families, since both models \ch{were} of a scaling type. Visible closeness \ch{in the left plot of Fig.~\ref{figitamada_nf_vs_ne}} of the experimental data points to model $A$ between $N_e\simeq (1-8)\cdot 10^6$  is most likely chance coincidence, and accelerated growth of the experimental $\sum{E_\gamma}$ there may be explained by a drop of the slope of gamma ray spectra, with respective increase of the average energy of families. In such a case it should be expected  that the experimental data rather correspond to the model $B$ everywhere, and there is superposition of some additional penetrative component, presumably protons, which results in local rise of the mean energy \ch{of gamma families in} the $N_e\simeq (1-8)\cdot 10^6$ interval, \idest between the primary energy limits of $3$\,PeV and $20$\,PeV.

In the right plot of Fig.~\ref{figitamada_nf_vs_ne} it is shown a set of dependencies $\sum{E_\gamma}$ on $N_e$ which was obtained in \ch{the} simulation \cite{tamada_simul_1994_e0_from_ne} for different primaries: protons, the nuclei of CNO group, and iron. It is seen there that using the curve for Fe nuclei as a basis for the $B$ model and superimposing over it the data for proton primaries between $N_e=(1-8)\cdot 10^6$ one could formally reproduce the experimental dependence. The drop of $\sum{E_\gamma}$ at $N_e\simeq 10^7$ could be explained by \ch{restoration of the previous relative share} of the heavy component in cosmic ray composition. Thus, the experimental data can be interpreted by a model which assumes a purely iron content of primaries at $E_0\simeq 3$\,PeV, and \ch{local} predominance of protons at $20$\,PeV.

Sudden appearance of penetrative primaries capable to carry larger part of its initial energy through the atmosphere agrees as well with observation, in the same energy range, of the long-flying cosmic ray component in \ch{the} former high altitude experiments at Tien Shan and Pamir \cite{longflying1,ontienlongfly2008,ontienlongfly2006,longflying_pamir2}.

With \ch{assumed hypothesis on existence of additional proton flux above the $3$\,PeV border} it remains the problem of two separate mechanisms \ch{which would ensure} particles acceleration up to \ch{two different thresholds of cut-off rigidity, $R_c\simeq 0.1$\,PV and $R_c\simeq 20$\,PV}. As an example of proper interpretation, it can be mentioned the model of a single close source of cosmic rays \cite{erlykin_wolfendale_1997}. In \cite{sveshnikova_close_source_2004} it was shown that the knee of the cosmic ray spectrum in that case remains at the same  3\,PeV point on the energy scale, and in \cite{sveshnikova_close_source_2013} that the deposit of such source can reveal itself as a local bump in the spectrum.

\ch{
A simpler, but exotic, hypothesis concerning the nature of the change in cosmic ray composition which reveals as a hardening of the secondary particles spectra in hadronic interactions is connected with suppositional existence of some non-nuclear cosmic ray component with specific interaction properties. As such the particles of strange quark matter \cite{centauro_bjorken,muon_stranglets_wilk-wlodarczyk_2017} might be proper candidate since their maximal charge $Z\sim 1000$ could make compatible existence of a single cut-off rigidity threshold at $R_c\simeq 0.1$\,PV with continuation of the cosmic ray spectrum above the $3$\,PeV border due to the common Fermi acceleration mechanism. In this connection it should be mentioned that the results gained up to date at Tien Shan
% in the experiments  \cite{shaulov_stranglets_1996} and HORIZON-T \cite{ontienhorizont_2019}
do not exclude their interpretation by a lost of} \ch{stability of the hypothetical particles of strange quark matter at their interaction with the air nuclei \cite{centauro_sqm,hadron_high_muons,shaulov_sqm_2019}.
}

It is easy to see that the discussed result of the high-altitude \HADRON experiment \ch{about domination} of the light  \ch{primaries} between $(3-20)$\,PeV is quite opposite to what was later concluded on composition of  cosmic rays in \ch{some} publications of the \KASCADE collaboration, \textit{e.\,g.} \cite{kascade_composition_2009,kascade_composition_at_icrc_2021}. According to their analysis,
% of the EAS data of a sea level experiment which strongly depends on concrete models of particles interaction,
it is the light component of cosmic rays that starts to die away at $E_0\simeq (3-5)$\,PeV, whereas the spectrum of heavy primaries \ch{(Fe)} lasts essentially further and prevails at a few tens of PeV. \ch{In contrary, there is a number of experimental evidences as well on predominance of protons in the same energy range around $(10-100)$\,PeV of the cosmic rays spectrum \cite{kascade_composition_2005,icetop_composition_2020}.}

\ch{
Mutual disagreement between the conclusions of various research groups concerning the composition of cosmic ray primaries may arise from diversity of information types obtained and analyzed in those experiments. One of such reasons is inequality of the atmosphere target in experimental sites located at the sea level and in mountains. Another one is the difference in the two investigation approaches based on the study of the hadronic and electromagnetic components of EAS: while an EAS core is directly formed due to the high-energy nuclear interactions, their role in producing of the electromagnetic part of a shower is largely veiled by cascade processes, particularly at low altitudes. As demonstrated by existing results of different experiments, in the second case the restoration task of primary composition is much more difficult and depends on parameters of the nuclear-electromagnetic cascade model applied for analysis of experimental data. In practice any simulation of EAS development demands some simplification of the model which decreases its predicting capability.
}

\ch{
The controversy revealed between the results concerning the average atomic mass of the PeV energy cosmic rays indicates direction for further studies, and should be finally resolved in the way of experiment.
%
% Possible reason for such controversy may be both the inadequacy of the models used for the analysis of experimental data, and the difference in the location height of the experimental installations above the sea level, as well as the fact that most modern experiments did not have any large-sized central part with a continuous, or sufficiently dense, arrangement of particles detectors which is utterly necessary for investigation of the high energy particles interaction in the core region of EAS.
%
It should be stressed also that this contradiction is of principal character, since it is the composition of primary particles that any possible hypothesis about the origin of the knee in the cosmic rays spectrum depends on.
}

\section{Conclusion}

The main results of the \HADRON experiment considered in this article can be summarized as the following.
\begin{itemize}

\item
Combination of an XREC, for individual detection of the most energetic had\-rons in the core of extensive air showers, and of the system of shower particles detectors, for determination of \ch{the} EAS parameters, made the hybrid experimental installation \HADRON an unique instrument for precise study of high energy hadronic interactions in cosmic rays. For the first time in this experiment the dependence of energy spectra of secondary EAS hadrons on the primary interaction energy $E_0$ was measured directly, \idest without any apriori supposition on the model of hadron interaction.

\item
It was found that the spectra of secondary hadrons born in interaction of cosmic ray particles can not be represented by some universal dependence in scaling manner over the whole range of investigated primary energies, $E_0\simeq (1-100)$\,PeV. Actually, the slope of the power energy spectrum diminishes essentially, and the average energy of hadronic products, correspondingly, increases between the values of $E_0\simeq (3-20)$\,PeV.

\item
At the same time, the modern accelerator experiments did not detect any deviation from scaling behavior in collisions of protons and nuclei in the same energy range \ch{of several PeV} (in the target rest frame). Hence, the effect of scaling violation found in cosmic rays must be connected with some non-homogeneity of their composition, and there must exist some specific cosmic ray component which could be responsible for detected hardening of the spectra of secondary hadrons in a limited range \ch{of primary energy} $E_0$.

\item
The energy threshold of $E_0\simeq (3-7)$\,PeV at which the said modification of \ch{the} secondary particles spectra in hadron interactions starts to reveal itself \ch{is close to} the position of the well-known knee in the \ch{energy spectrum of primary} cosmic rays. Since purely chance coincidence of such heterogeneous effects seems to be rather improbable, it \ch{might} be supposed that both phenomena should be stipulated by one and the same physical reason.

\end{itemize}

\ch{
Thus, possible explanation of the effect of scaling violation detected by the direct investigation of high-energy cosmic rays in the \HADRON experiment presently assumes two alternative possibilities. These are either an additional flux of protons generated by nearby sources in vicinity to the Solar system, or an exotic variant of some non-nuclear component with specific interaction properties which appears among cosmic rays at the energy of a few PeV. The nearest plans of the experimental activity at the Tien Shan mountain station anticipate continuation of the complex study of particles interaction processes which take place in the knee region of the cosmic ray spectrum, with general purpose to elucidate the probability of mentioned alternatives.
}

\section*{Acknowledgements}

The \HADRON experiment was accomplished due to efforts of a large team of participants from LPI. The authors are truly thankful to all of them for their work. One of the authors thanks also his colleagues from the international experiment PAMIR with whom he collaborated during many years by mastering the method of X-ray emulsion films.

This study was supported by the RFFI grant \#98-02-16942 and by the grant \#AP08955730 of the Ministry of Education and Science of the Republic of Kazakhstan.

\section*{References} % this must be commented out!
%
% BibTeX users please use
%\bibliographystyle{elsarticle-num} %{unsrt}
%\bibliography{p,cr,thisherehadron}
%\end{document}

\newcommand {\etal}{\textit{et al.}, }

\end{document}